\documentclass{aa}
 \usepackage{graphicx}
  \begin{document}

 %\thesaurus{03(....... 3C279)}

 \title{Connection between optical and radio/millimeter flares in blazar OJ287}
 %\volnopage{}
 %\setcounter{page}{1} 
 \author{S.J.~Qian\inst{1}}
 \institute{National Astronomical Observatories,
    Chinese Academy of Sciences, Beijing 100012, China} 
  %\and Department of
  % Astronomy (IAG-USP), University of Sao Paulo, Sao Paulo 05508-090, Brazil}
  % \and UCLA, Department of Physics and Astronomy, LA, CA90095, USA}       
 \date{Complied by using A\&A latex}
  \abstract{Blazar OJ287 is a unique source in which optical 
  outbursts with double-peak structure have been observed  quasi-periodically
   with a cycle of $\sim$12\,yr. It may be one of the best candidates for 
   searching supermassive black hole binaries.}{We investigate the connection
    between its optical and radio/millimeter variations and interpret the 
  emissions in terms of relativistic jet models.}{Specifically, we make a 
   detailed analysis and model simulation of the optical and radio/mm
   light curves for the outburst  during the period  of 1995.8--1996.1.}
     {It is shown that the multi-wavelength light-curves at optical V-band and
    radio/mm wavelengths (37, 22, 14.5 and 8\,GHz) can be decomposed into
    36 individual elementary flares, each of which have symmetric 
   profiles.}{The elementary flares can be understood to be produced through
   lighthouse effect due to the helical motion of  corresponding
   superluminal optical/radio knots. Helical motion of superluminal
   knots should be prevailing in the inner regions of its relativistic jet 
   formed in the magnetosphere of the putative supermassive black 
   hole/accretion disk system. A comprehensive and compatible framework 
   for understanding the entire phenomena in OJ287 is described.}
  \keywords{galaxies: active -- galaxies: jets --
  galaxies: nuclei -- galaxies : individual OJ287}
  \maketitle
  \section{Introduction}
   OJ287 (z=0.306) is  one of the best studied and most prominent blazars.
   It has shown large amplitude and rapid variability at all wavelengths
   from radio, optical through to $\gamma$-rays with various time-scales 
   from hours to years. A large amount of observations in radio, infrared, 
   optical, ultraviolet, X-rays and $\gamma$-rays have been carried out
    to investigate the
   nature of the phenomena in this source (e.g., Usher \cite{Us79},
    Holmes et al. \cite{Holm84}, Brown et al. \cite{Bro89}, Hartman et al.
    \cite{Har99},  Sillanp\"a\"a  et al. \cite{Si88}, \cite{Si96a},
   \cite{Si96b},  Takalo \cite{Tak96a}, Takalo et al. \cite{Tak96b}, 
   Allen et al. \cite{Allen82}, Aller et al. \cite{Al94}, \cite{Al14},
   Tateyama et al. \cite{Ta99}, Valtaoja et al. \cite{Val20},
    Qian \cite{QiXiv18}, Britzen et al. \cite{Br18}, Hodgson et al.
   \cite{Hod17}, Valtonen et al. \cite{Va17}, Cohen \cite{Co17}, Kushwaha 
   et al. \cite{Ku18a}, \cite{Ku18b}, Gupta et al. \cite{Gu16}, 
   Agudo et al. \cite{Ag12}, Ackermann et al. \cite{Ac11}, Neronov \& Vovk
   \cite{Ne11}).\\
    With the distinct properties common for generic blazars, 
   the most remarkable characteristic of OJ287 is the quasi-periodic 
   variability in
   optical wavebands. The optical light curve
   recorded since 1890s reveals quasi-periodic outbursts 
   with a cycle of $\sim$12\,yr (Sillanp\"a\"a et al. \cite{Si88}). 
   Up to now four quasi-periodic outbursts with double-peak structure
   have been observed in 1971--73, 1983--84, 1994--95 and 2005--2007.
   The first flare of the fifth quasi-periodic outburst was observed
    in December 2015 and the second one has been predicted to appear soon
    (Valtonen et al. \cite{Va18}, Dey et al. \cite{De18}).\\
   The quasi-periodicity in the optical outbursts and their double structure
     are widely believed to be related to the orbital motion of
    the putative black hole binary in its nucleus. The very complex
    variations in flux density and polarization at all 
    wavelengths from radio to $\gamma$-rays are mostly believed to occur in 
    the relativistic jet of the primary hole (or both jets produced 
    by the primary and secondary holes; Qian \cite{QiXiv18}).
     Various theoretical models (or scenarios) have been proposed to interpret
    the whole phenomena observed in OJ287 (for details, referring to the 
    discussions in Villforth et al. \cite{Vil10}  and  Qian \cite{QiXiv18},
    \cite{Qi15}). On the whole, these models can be divided into two 
   categories: disk-impact models and relativistic jet models, both involving
   a supermassive black hole binary in the nucleus of OJ287.
    \begin{itemize}
   \item   The disk-impact model originally proposed by Lehto \& Valtonen
   (\cite{Le96}) has been steadily improved (Valtonen \cite{Va07},
    Valtonen et al. \cite{Va06}, \cite{Va18},
     Dey et al. \cite{De18}) to interpret the double-peak structure of the
   quasi-periodic optical outbursts. It suggests that 
   the first major flares of the quasi-periodic outbursts are produced by 
   the secondary hole penetrating into the accretion disk of the primary hole
   and are thus bremsstrahlung in origin. In the case of highly 
   eccentric orbital motion with large inclination, two impacts would 
   occur per pericenter passage, causing the double-peak structure. The
   follow-up flares, as well as the non-periodic outbursts occurred during 
   the intervening periods are suggested to be related to the enhanced 
   accretion events induced by the disk-impacts and tidal interaction 
   between the
   secondary hole and the primary disk. This model requires
   a high inclination angle ($\rm{i}$\,${\sim}$$50^{\circ}$\,--\,$90^{\circ}$)
   and a large orbit eccentricity ($\rm{e}$\,$\sim$\,0.66) and
   a strong constraint on the total mass of the binary, reaching 
   ${\sim}2{\times}{10^{10}}{M_{\odot}}$ with a mass ratio m/M$\sim$0.007.  
   This model has been elaborated to make accurate timing of the quasi-periodic
   outbursts. \\
   The cavity-accretion flare model proposed by  
   Tanaka (\cite{Tan13}) 
   is another type of disk-impact model, which assumes that the primary hole
   and the secondary hole having comparable masses and are in near-coplanar
   orbital motion. Hydrodynamic/magnetohydrodynamic (HD/MHD) 
   simulations for such binaries surrounded by circumbinary disks
    have shown that cavity-accretion processes would create 
   two gas-flow streams impacting onto the disks of the black holes per
    pericenter passages, possibly causing the double-peak structure of the 
   quasi-periodic outbursts. This model also suggest that the gas-flow impacts
   produce thermal outbursts, but it is not able to make
    accurate timing of the quasi-periodic outbursts. The follow-up flares 
   and the non-periodic outbursts during the intervening periods were 
   not investigated in this model.  
   \item It is widely suggested that blazars are extragalactic sources with
   relativistic jets pointing close to our line of sight, and they are the 
   brightest and most violently variable active galactic nuclei in radio,
   optical through to $\gamma$-rays. Relativistic jet models have been applied
   to interpret the optical and radio variability behavior in OJ287  by many 
   authors (e.g., Sillanp\"a\"a et al. \cite{Si96a},
   Valtaoja et al. \cite{Val20}, Villata et al. \cite{Villa98}, 
   Villforth et al. \cite{Vil10}, Britzen et al. \cite{Br18},
   Qian \cite{Qi15}, \cite{QiXiv18}). \\
   Recently, Qian (\cite{QiXiv19}) has
   tentatively proposed an alternative jet model to understand the entire
   phenomena in OJ287, which is based on the optical  multi-wavelength 
    and $\gamma$-ray observations performed for OJ287 
   (e.g., Kushwaha et el. \cite{Ku18a}, \cite{Ku18b}), combining  with 
   the distinct features previously found in the optical and radio 
   (flux and polarization) variations (e.g, Sillanp\"a\"a et al. \cite{Si96a},
   Valtaoja et al. \cite{Val20}, Usher \cite{Us79}, Holmes et al. 
    \cite{Holm84}, Kikuchi et al. \cite{Ki88}, D'Arcangelo et al. \cite{Da09},
    Kushwaha et al. \cite{Ku18a}, Britzen et al. \cite{Br18}, 
   Qian \cite{QiXiv18}).\\
    The main emission properties are explained, including 
    the following aspects (for details, referring
   to Qian \cite{QiXiv19}). (1) The $\gamma$-ray flare observed in 
   December/2015 was observed to be 
   closely associated with the optical outburst. The simultaneity between
   the optical and $\gamma$-ray flares implies that the optical outburst in
   December/2015 (peaking at 2015.92) should be produced in the jet and 
   a nonthermal (synchrotron) flare;
   (2) The optical multi-wavelength observations (Kushwaha et al.
    \cite{Ku18a}) have shown that the December/2015 optical outburst has its 
   temporal and spectral variability behavior very similar to that of the
   synchrotron outburst in March/2016. The two outbursts form a pair of flaring
   events with a time-interval of $\sim$90\,days. Thus the December/2015 
   outburst must
   be a synchrotron flare and can not be a thermal flare as claimed in the
   disk-impact model;
   (3) Previous observations have already shown that the major optical
    outbursts can be  decomposed into a number of elementary flares, each
    having a symmetric flux density profile. Symmetry in the light-curves
    has been recognized as a distinct feature of the optical flares in OJ287;
   (4) The simultaneous variations in radio/mm and optical bands without 
   measurable time-delays ($<$1-2\,days) and the similarity between
    the envelopes of optical and radio/mm outbursts  imply that these 
   correlations can only be interpreted in terms of lighthouse effect due to
   the helical motion of superluminal optical/radio knots;
    (5) Under the lighthouse scenario the December/2015 and the March/2016
    outbursts can be interpreted as produced when the superluminal knot (shock)
    propagates along the helical magnetic field in a perfect-collimation zone
    through two revolutions.
   \end{itemize}
    As shown in Qian (\cite{QiXiv19})
   The light-curves of six quasi-periodic optical outbursts (in 1983.0, 1984.1,
   1994.8, 2005.8, 2007.7 and 2015.9)
   and a few  individual non-periodic outbursts (in 1993.9, 1994.2 and 2016.98)
   have been model-simulated and well interpreted in terms of the lighthouse
    model. We suggest  that all these outbursts are 
    synchrotron flares produced in the relativistic jet.\\
    While this relativistic jet model can well explain the light-curves of both
   the periodic and non-periodic outbursts,\footnote{Relativistic jet models
    are usually applied to explain the flaring properties in generic blazars
   from radio, optical (flux and polarization) through to $\gamma$-rays.} 
   it needs to invoke the dual-stream
   flow accretion mechanism to interpret the double-structure of the 
   quasi-periodic  outbursts as in the cavity-accretion flare model by Tanaka
   (\cite{Tan13}). In principle, this is possible as some MHD simulations
   have demonstrated (e.g., Artymovicz \& Lubow \cite{Ar96},
    Artymoviz \cite{Ar98}, Hayasaki et al. \cite{Ha08}, Shi et al. 
   \cite{Sh12}, D'Orazio et al. \cite{Do13}), but detailed modeling is 
   imperatively required, especially for the timing of the optical outbursts
     (the $\sim$12\,yr quasi-periodicity and the $\sim$1-2\,yr
    time-separation  of the double-peak structure). Simulations of the
    accretion events causing all the outbursts during the whole 12\,yr period
    (or even during the 120\,yr period) may be also important.\\
   We point out that the relativistic jet models distinguish from the
    disk-impact models mainly in two aspects.\\
    (1) In the relativistic jet models all the optical outbursts (both 
    quasi-periodic and non-periodic) originate from synchrotron process and 
    thus are related to its primary-hole jet (or to both jets of the primary 
    and secondary holes, Qian \cite{QiXiv18}).
    In the disk-impact models the primary quasi-periodic optical
     outbursts are assumed to 
    originate from bremsstrahlung process due to the secondary hole impacting 
    onto the primary-hole disk, while the follow-up and non-periodic
    outbursts are assumed to originate from synchrotron process, 
    related to its primary hole jet.\\ 
    (2) Therefore, in the relativistic jet models, the energetics of the 
    optical outbursts is physically 
    unified: the time-scale of all  the optical outbursts is 
    compressed and their flux density boosted by Doppler effects in the same
     way. The light-curve of all the optical outbursts can be considered 
    in a unified 
    physical frame. In contrast, the disk-impact models require dual-energetics
    : the synchrotron outbursts are Doppler boosted while the impact 
    (thermal) outbursts are not. This implies that one needs different
    scales (both time-scale and flux density scale) to describe their 
    energy content and scaling of light-curve for the two types of outbursts.
    For example, a synchrotron optical flare could have been Doppler boosted
    by a  factor of $\sim{10^5}$ (see Section 3), corresponding to a brightness 
    magnification of $\sim$12.5\,mag, while the impact flares are not 
    Doppler-boosted.\\
     In this paper we will investigate the simultaneous radio/millimeter and
    optical variations occurred in OJ287 and provide more evidence for the
    validity of the lighthouse scenario proposed in Qian (\cite{QiXiv19}).
     \begin{figure*}
    \centering
    \includegraphics[width=6cm,angle=-90]{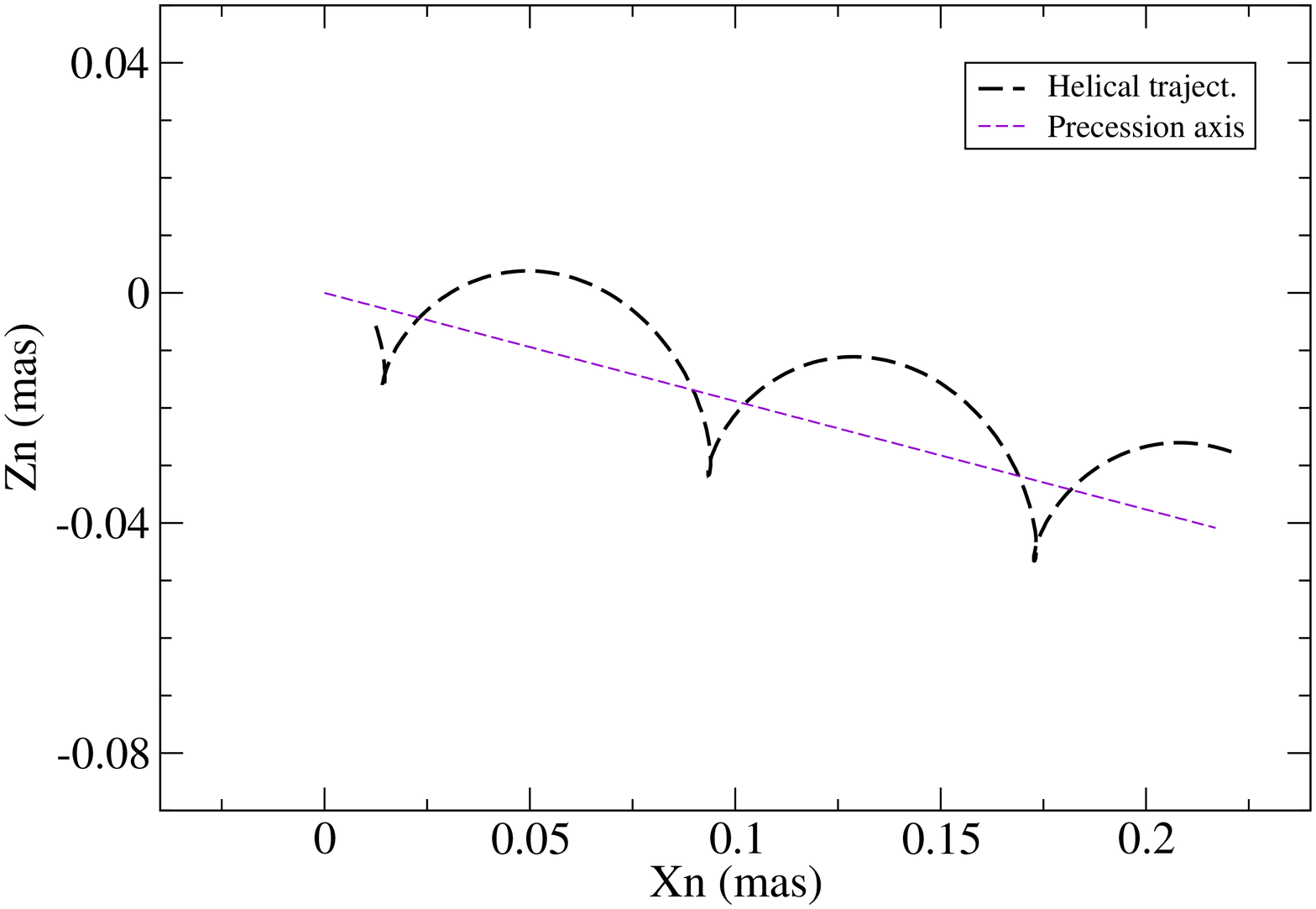}
    \includegraphics[width=6cm,angle=-90]{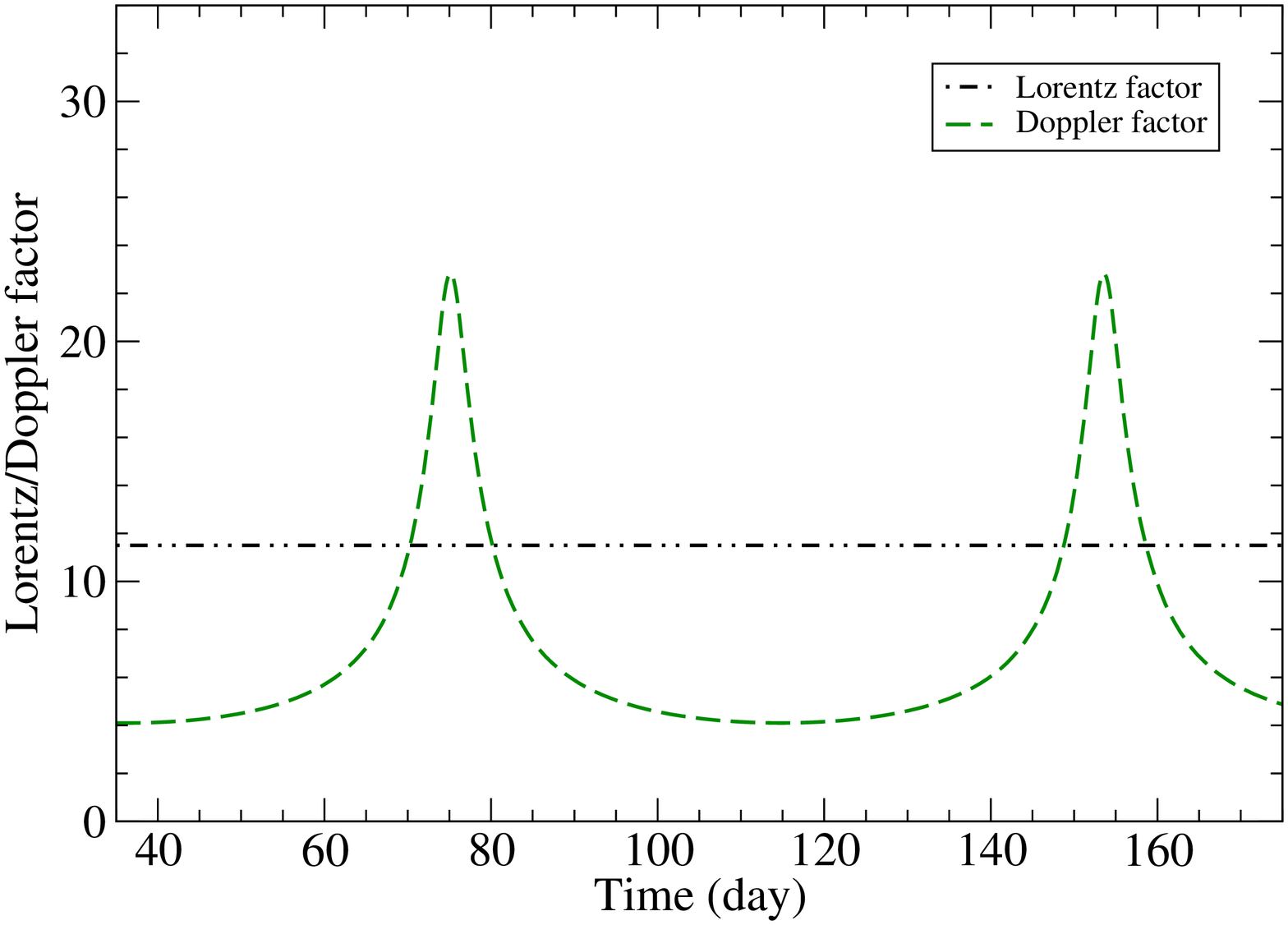}
    \caption{Left panel: Scenario for a precessing jet and helical motion. 
    The straight line denotes the precessing 
    jet axis (projected on the plane of the sky) which is defined by 
    precession phase $\omega$=--2.0\,rad. The helix indicates the trajectory 
    of the optical knot moving around the jet axis in the perfect collimation
     zone for the elementary flare in 1995.98 (peaking at JD2450075). The
    corresponding Lorentz and Doppler factors are shown in the right panel.}
    \end{figure*}
    \begin{table}
    \caption{Parameters defining the precessing jet-nozzle scenario and the
     helical trajectory of the superluminal knot, which are used in the model
    simulation of the optical and radio light-curves.}
    \begin{flushleft}
    \centering
    \begin{tabular}{ll}
    \hline
    Parameter & fixed value \\
    \hline
    $\epsilon$ & $3^{\circ}$ \\
    $\psi$  & 0.0\,rad \\
    $\omega$ & --2.0\,rad \\
    $a$   & 0.0402 \\
    $x$   &  1.0 \\
    $A_0$ & 0.0138\,mas \\
    d$\phi$/d$z_0$ & -7.04\,rad/mas \\
    \hline
   \end{tabular}
   \end{flushleft}
   \end{table}
   \begin{table}
   \caption{Base-level (or underlying jet) spectrum selected for the
    1995.84 optical and radio/mm outbursts.}
   \begin{flushleft}
   \centering
   \begin{tabular}{ll}
   \hline
   Waveband & Flux density \\
   \hline
   37\,GHz & 2.31\,Jy \\
   22\,GHz & 2.31\,Jy \\
   14.5\,GHz  & 1.30\,Jy \\
   8\,GHZ &  1.30\,Jy \\
   V-band & 2.70\,mJy \\
   \hline
   \end{tabular}
   \end{flushleft}
   \end{table}
   \begin{table}
   \caption{Modeled spectrum between 8\,GHz and V-band
    of the simultaneous radio-optical flare observed in 1995.98 (peaking at 
    JD2450075). The K-, U- and UV-band flux densities are given for comparison.
    They are derived from the modeled V-band flux density, assuming a 
    broken power-law spectrum at the infrared-optical-UV wavelengths 
    with the spectral index changing from 0.8 to 1.3 at the V-band break
    (Qian \cite{QiXiv19}, Kushwaha et al. \cite{Ku18a}).}
   \begin{flushleft}
   \centering
   \begin{tabular}{ll}
   \hline
   Waveband & Flux density\\
   \hline
   37\,GHz & 1.30\,Jy \\
   22\,GHz & 1.37\,Jy \\
   14.5\,GHz & 1.52\,Jy \\
   8\,GHz &   1.44\,Jy \\
   K-band & 29.3\,mJy \\
   V-band &  9.59\,mJy \\
   U-band & 5.62\,mJy \\
   UV-band & 1.06\,mJy \\
   \hline
   \end{tabular}
   \end{flushleft}
   \end{table}
   \begin{table}
   \caption{Modeled comoving (or de-boosted) flux densities for the
     elementary flare in 1995.98 (peaking at JD2450075). Numbers 
     in the parentheses represent the powers of ten.}
   \begin{flushleft}
   \centering
   \begin{tabular}{ll}
   \hline
   Waveband & Flux density \\
   \hline
   37\,GHz &  4.79(-6)\,Jy \\
   22\,GHz & 5.05(-6)\,Jy \\
   14.5\,GHz & 5.61(-6)\,Jy \\
   8\,GHz & 5.31(-6)\,Jy \\
   V-band &  3.53(-5)\,mJy\\
   \hline
   \end{tabular}
   \end{flushleft}
   \end{table}
    \begin{figure*}
    \centering
    \includegraphics[width=6cm,angle=-90]{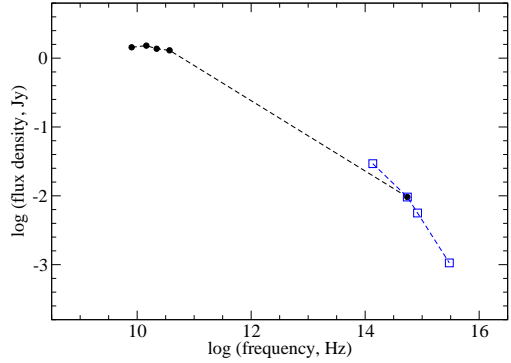}
    \caption{Modeled spectrum of the elementary optical/radio flare in 1995.98
    (peaking at JD2450075; filled black circles). In order to simulate
   the radio light-curve the spectral indexes between the optical and 
   radio frequencies  are selected to be: 
    ${\alpha}_{37,\rm{v}}$$\equiv$${\alpha}$(37GHz,V-band)=0.51,
    ${\alpha}_{22,\rm{v}}$$\equiv$${\alpha}$(22GHz,V-band)=0.49,
    ${\alpha}_{15,\rm{v}}$$\equiv$${\alpha}$(14.5GHz,V-band)=0.48 and
     ${\alpha}_{8,\rm{v}}$$\equiv{\alpha}$(8GHz,V-band)=0.45,
   respectively. The K-, U- and UV-band flux densities (open squares) are
   displayed for comparison. No data are available for the wavelengths 
   between 37\,GHz and V-band and the dashed line is only 
   as an aid to the eye.}
    \end{figure*}
 \section{Basic considerations}
   The variability behavior of blazar OJ287 is very complex. In order to
    understand the entire phenomena observed in this blazar,  a detailed 
    summary analysis of its flaring characteristics observed at radio,
   infrared, optical wavelengths and in $\gamma$-rays have been made in Qian 
   (\cite{QiXiv19}), including: (1) simultaneous optical and $\gamma$-ray
    flares; (2) spectral energy distribution of the outbursts; (3) similarity
    in temporal and spectral variations of quasi-periodic outbursts 
    and non-periodic (synchrotron) outbursts;
    (4) simultaneity in radio/millimeter and optical variations; 
    (5) color stability during the NIR-optical-UV outbursts (for both
    quasi-periodic and non-periodic) ; (6) broken power-law spectra in 
   NIR-optical-UV bands, (7) decomposition of light-curves of optical
   outbursts into elementary flares;  (8) symmetry in the profiles of optical 
   outbursts; (9) possible existence of double jets produced by the putative
   black hole binary system; (10) connection between the optical 
   outbursts and the
   emergence of superluminal radio knots; (11) distinctive variations in
   optical and radio polarization; (12) similarity in light-curve structure of
   the quasi-periodic outbursts (e.g. Sillanp\"a\"a et al. \cite{Si85}); (13)
    association of low polarization degree with rapid polarization angle 
   change in quasi-periodic optical outbursts (e.g., Villforth et al. 
   \cite{Vil10}); etc.\\
    Based on the summarization  of all these characteristics, the 
    precessing jet nozzle model previously proposed by Qian et al.
    (\cite{Qi91}, \cite{Qi17}, \cite{QiXiv18}) has been used to simulate
     the light-curves of the quasi-periodic outbursts in 1983.00, 1984.10,
     1994.75, 2005.76, 2007.70, 2015.75 and a few non-periodic flares.
    It has shown that both periodic and non-periodic optical outbursts
    may be synchrotron flares originated from the relativistic jet. Their 
    light curves can be interpreted in terms of lighthouse effect due to the
    helical motion of the superluminal optical knots.\\
    Here we would like to mention some results of the investigations 
    obtained in earlier years as supplementary arguments for the synchrotron 
   origin of the quasi-periodic optical outbursts in OJ287.
    \begin{itemize}
     \item Usher (\cite{Us79}) studied the multi-frequency light-curves
    of the outburst observed in OJ287 during the period of 1971--1975 
    \footnote{The light-curves were compiled from the data published in
      literature at wavelengths of
    0.44$\mu$m, 10$\mu$m, 0.35, 0.9, 2.8, 3.8, 4.5, 6, 11, and 18\,cm.} and
    suggested that there is a multi-frequency optical-radio synchrony 
    within observational uncertainty for bursts III, IV and V 
    (as called in that paper), which peaked 
    at $\sim$1973.0, 1973.9 and 1975.1 respectively. Burst II  
    (peaking at $\sim$1971.7) could be also synchronous at optical and radio
     wavelengths, if it is the result of superposition of two or more bursts. 
    It is noted that optical bursts II and III (the strongest and secondary
    strong flares in this outburst event) have been identified as the double
    ``impact (thermal) flares'' of the quasi-periodic outburst 
    (Dey et al. \cite{De18}). Thus the radio variations synchronous with the
     optical variations observed for bursts II and III may imply
    that the optical outbursts are nonthermal (synchrotron) flares. 
    In this paper  we will investigate the correlation between the optical 
    and radio variability observed in the  quasi-periodic outburst during 
    1995.8--1996.1 in OJ287 and further show that the radio variations 
    simultaneous with the optical variations indicate that this optical 
    outburst is nonthermal.\\
     \item Holmes et al. (\cite{Holm84}) made simultaneous 
     observations of OJ287 at infrared JHK wavelengths and in optical and near
     infrared wavelengths at UBVRI in 1983 January (during 4 days from 
     January 7 to January 10). This time-interval just coincided with the 
     peaking stage of the primary
     quasi-periodic outburst (Valtaoja et al. \cite{Val20}), which was 
     identified as an ``impact (thermal) burst'' in the precessing binary model 
     (Lehto \& Valtonen \cite{Le96}, Sundelius et al. \cite{Su97}, 
     Valtonen et al \cite{Va18},
      Dey et al. \cite{De18}).\footnote{The starting epoch of the 1983.0 
     outburst (1982.964) has been taken as a standard reference-epoch for
     the outburst timing in all versions of the disk-impact model 
    (Lehto \& Valtonen 
     \cite{Le96}, Sundelius et al. \cite{Su97}, Valtonen et al. \cite{Va18}).}  
    The very low polarization degrees of p$<$1\%
     at R- and V-bands observed on January 8 (a dip in the polarization curves)
     were regarded as the rational evidence for the optical
     outburst being a thermal flare. However, the rapid daily variations in
     polarization angle at R-band ($\sim{60^{\circ}}$) and V-band 
     ($\sim{110^{\circ}}$) during 
    one-day period (from January 8 to January 9)\footnote{Similar polarization
    behavior was observed at B-band and U-band.} does not
     support this  interpretation, because appearance of a thermal outburst
    (with zero polarization)
     can greatly reduce the polarization degree of the source, but can not
     cause its variations in polarization angle. This argument seems 
    important: occurrence of  very low polarization
     degrees ($<$2-4\%) alone cannot be simply recognized as a sign of
     strong thermal outburst appearing, because the associated variations
     in polarization angle must be due to synchrotron components.\\
      In order to interpret the simultaneous multi-wavelength light-curves 
    of time-variable flux density S($\nu$,t), polarization degree p($\nu$,t) 
   and polarization angle PA($\nu$,t)
    in K-, H-, J-, I-, R-, V-, B-, and U-bands, Holmes et al. proposed a 
    two-component model (also see Bj\"ornsson \cite{Bj82}, Bj\"ornsson \&
    Blumenthal \cite{BjB82}, K\"onigl \& Choudhuri \cite{Ko85} for
     jet polarization models).
    \footnote{It should be noted that this model seems to be one of the most
     effective jet models to coherently explain the complete 
    characteristics of the optical multi-wavelength flux density and 
    polarization for OJ287.}
     They strongly suggested  that the two
    components are physically connected with one component having a stable 
    polarization angle while the other one showing a gradual rotation. This
    behavior may be understood in terms of a physical rotation of the 
    magnetic field. Obviously, their modeling results are consistent with the 
     helical motion model (lighthouse model) proposed in Qian 
    (\cite{QiXiv19}) and the two model-components might be related to the
    jets produced by the primary and secondary black holes, respectively 
    (Qian \cite{QiXiv18}). In addition, Kikuchi et al. (\cite{Ki88}) observed 
    a synchronous rapid variations of polarization angles during February 3--12
    1986 in the optical and radio wavebands on time-scale of days. They 
    explained this correlated variation in terms of  a shock of 
    ``core-envelope'' structure moving along the helical magnetic field 
    in the jet. Moreover, the 1983.0 outburst has spectral curvatures 
    in infrared-optical and optical-ultraviolet regions 
    with concave or convex structures which can be also explained in terms
    of the two-component model together with the polarization features.
    It seems that the distinct spectral structures observed for the 2015 
    quasi-periodic outburst in IR-optical-UV wavebands (Kushwaha et al.
    \cite{Ku18a}) could be interpreted in terms of a jet model like that 
    proposed by Holmes et al. (\cite{Holm84}).\\
     We further point out that very low polarization degrees were observed
    in the quasi-periodic optical outbursts in September 2007 (Villforth et al.
    \cite{Vil10}) and December 2015 (Valtonen et al. \cite{Va17}). In both
    cases the associated rapid variations in polarization angle 
    may indicate that the optical outbursts are nonthermal flares produced
    in the relativistic jet. The time-variable polarization degree curve should
    be interpreted coherently with the interpretation of the time-variable 
    flux density curve and polarization angle curve. 
    \item D'Arcangelo et al. (\cite{Da09}) explored the correlation between 
    polarization characteristics at optical and near-infrared (NIR)
    wavelengths and 
    that in the 43\,GHz compact core in OJ287 between 2005 October 24 and 
    November 3  and between 2006 March 27 and April 4. They found that during
    this period the polarization degrees in the optical, near-infrared 
    and radio-core are nearly steady. They measured the mean R-band 
    polarization degree to be 30.7\% which appears to oscillate with a 
    period of $\sim$ 2\,days. Interestingly, the polarization
    degrees in B-, V-, R- and I-bands have similar values and synchronous 
    variations. D'Arcangelo et al. fit the optical spectra with 
    a power-law and found an steady optical spectral index of 
     $\sim$1.28$\pm$0.05.  We notice that the first (strongest) flare 
    of the quasi-periodic optical outburst in 2005.74, which was identified as
    a thermal impact-flare in the  precessing binary 
    model (Valtonen \& Ciprini \cite{Va12}), peaked between
    2005 October 17 and October 28. Thus the high polarization degrees 
   ($\sim$30\%) observed at multi-wavelengths (B-, V-, R- and I-wavebands) 
    by D'Arcangelo et al. between October 25 and 28 should be ascribed to 
    this optical flare, demonstrating its origin of synchrotron radiation.
    The short rotation (from $20^{\circ}$ to $8^{\circ}$) measured between 
    2005 October 25 and October 29 also favors relativistic jet models
    (Qian \cite{QiXiv19}).\\
    Moreover, D'Arcangelo et al. suggested that the variable optical-NIR
    emission observed during the two campaigns in 2005 and 2006 
    originates from the 43\,GHz core region of the parsec-scale jet. They
    proposed a spine-sheath model to interpret the synchronous optical and
    radio polarization variability, including all the characteristics
    of the observed emission: the time-variable flux density 
    S($\nu$,t), polarization degree p($\nu$,t) and polarization angle  
    PA($\nu$,t).
     \end{itemize}
     The results restated above for the quasi-periodic optical outbursts in 
     1971.7, 1973.0, 1983.0 and 2005.7 show that they are all nonthermal 
    flares, produced within the relativistic jet, and support the lighthouse 
    model (or helical motion model) proposed in Qian (\cite{QiXiv19}).\\     
     It can be seen from the above description that multi-frequency observations
    in optical/IR and radio/millimeter wavebands (measuring time-variable 
    flux density curve, polarization degree curve and polarization angle curve)
    are very important to understand the nature of optical emission in OJ287.
    Search for correlation between the optical and radio outbursts can 
    provide decisive evidence for the mechanism producing the optical 
    radiation. In this paper we will investigate the optical-radio correlation
    for the quasi-periodic optical outburst event in 1995.8.\\ 
    We would like to point out that a variety of connections between 
     optical and radio variability has been observed in OJ287, involving
    different emitting processes: e.g., (1) 
    simultaneous radio-optical bursts; (2) radio bursts delayed with respect to
    optical bursts (e.g., Valtaoja et al. \cite{Val87}, \cite{Val20}); 
    (3) radio bursts without optical counterparts; (4)
    optical bursts without radio counterparts, etc.  Superluminal optical 
    knots, which produce simultaneous radio-optical emission through lighthouse 
    effect near the nucleus, may continue to move outward and evolve 
    to form superluminal radio knots 
    on parsec scales, causing delayed connection between radio and optical
    outbursts (e.g., Britzen et al.
    \cite{Br18}, Qian \cite{QiXiv18}, Agudo et al. \cite{Ag12}, Sillanp\"a\"a
    et al. \cite{Si85}, Valtaoja et al. \cite{Val20},
    Tateyama et al. \cite{Ta99}, Kushwaha et al. \cite{Ku18a}, Aller et al. 
    \cite{Al14}).
     \section{Assumptions}
     In the following, we will continue to explore the optical and radio 
    variability behavior in OJ287 in terms of the lighthouse model under the
    precessing jet nozzle scenario and concentrate on investigating the 
    connection between the radio/millimeter and optical flares observed 
    during the quasi-periodic outburst event in 1995.84, and 
    further show that the simultaneous variability behavior at radio/millimeter
    and optical wavelengths  support the lighthouse model proposed in
    Qian (\cite{QiXiv19}).
  \subsection{Geometric parameters}
   We will use the same formulation and parameters to describe the precessing
   jet-nozzle scenario and the helical model, referring to Qian
    (\cite{QiXiv19}) for  details. The geometric parameters for the model
   simulation are listed in Table 1. We assume that superluminal knots move
  along helical trajectories (probably helical magnetic fields) around the
   jet axis, which is shown in Figure 1 (left panel). The jet axis 
   precesses around the precession axis (not displayed here) which is defined 
   by parameters $\epsilon$ and $\psi$. The selected jet axis is defined
   by precession phase $\omega$=--2.0\,rad.
   The jet axis is described by parameters $a$ and $x$, which define a 
   straight line. The helical trajectory pattern is defined by the parameters
    $A$ (amplitude) and d$\phi$/d$z_0$ (rotation rate). In the right panel of
   Figure 1 the modeled Lorentz factor and Doppler factor are shown for the
   elementary flare in 1995.98 (peaking at JD2450075). 
   \subsection{Spectra of elementary flares}
    As suggested in Qian (\cite{QiXiv19}) that the quasi-periodic outbursts 
    can be decomposed into several elementary flares with symmetric profiles 
    and each elementary flare can be explained in terms of the lighthouse
    model. Thus the light-curves observed in optical and radio/mm wavebands
    can be simulated.  We assume that the elementary flares have broken 
   power-law spectra in the
   NIR--optical--UV bands with a spectral break at V-band of $\Delta{\alpha}$=
   0.5 and the local V-band spectral index ${\alpha}_{\rm{v}}$=1.0
    ($S_{\nu}$$\propto{{\nu}^{-{\alpha}}}$). The flux densities of the 
   underlying jet (or base-levels) are assumed to be constant and 
   listed in Table 2. The modeled
   radio/mm spectra are described in terms of the modeled indexes between
   the radio/mm and V-band. As an example, in Figure 2 is shown
   the modeled radio-optical spectrum for the elementary flare in 1995.98 
   (peaking at JD2450075). The modeled flux densities are listed in Table 3. 
   The corresponding comoving (or intrinsic) flux densities are listed 
    in Table 4.\\
    It should be noted that the local spectral index in the radio-mm region
   (8\,GHz to 37\,GHz) is much smaller than that at V-band 
   (${\alpha}_{\rm_{v}}$=1.0; see Figure 2),
    and thus the variability amplitudes of the radio/mm outbursts will be much 
   smaller than that in the optical V-band (see below).
      \subsection{Doppler boosting and simulation of optical light-curve}
  We will model simulate the V-band optical light-curve of the 1995.84 
  outburst event in terms of the lighthouse model proposed 
  in Qian (\cite{QiXiv19}).
  In this case  the lighthouse effect results in a symmetric light-curve via 
  Doppler boosting per revolution of the helical motion for each of elementary
  flares. The observed optical flux density can be written as:
   \begin{equation}
    {S_{\rm{v,obs}}}(t)={S_{\rm{v,co}}}
                  {\times}[{\delta}(t)]^{3+{\alpha}_{\rm{v}}}+{S_{\rm{b,v}}}
   \end{equation}
  Where $S_{\rm{v,co}}$(t) represents the constant comoving (or de-boosted) 
  optical flux density, ${\delta}$(t)--Doppler factor of 
   the superluminal optical knot,
   ${\alpha}_{\rm{v}}$--local spectral index at V-band 
   (Blandford \& K\"onigl \cite{Bl79}), assumed to be ${\alpha}_{\rm{v}}$=1.0.
   $S_{\rm{v,b}}$-- the underlying flux density of the jet at V-band (Table 2).
   The modeled Lorentz factor $\Gamma$, Doppler factor $\delta$ and comoving
   flux density $S_{\rm{v,co}}$ are selected for each of the elementary
    flares to   reasonably fit the entire optical light-curve (see Table 5).   
 \subsection{Parameters for simulation of radio/mm light-curves}
    For model simulation of the radio/millimeter light-curves at frequencies of
   8, 14.5, 22, and 37\,GHz, we will define the modeled two-point spectral
    indexes ${\alpha}$(radio,\rm{v}) between the radio/mm frequencies and 
   V-band frequency: ${\alpha}$(37GHz,\rm{v}),   ${\alpha}$(22GHz,\rm{v}),
    ${\alpha}$(15GHz,\rm{v}) 
   and ${\alpha}$(8GHz,\rm{v}).  The modeled radio/mm light-curves will be 
    derived from the modeled V-band light-curve
   using the formula given in equation (2).
   \begin{equation}
    {S_{\rm{radio}}}={S_{\rm{v}}}{\times}[{{\nu}_{\rm{radio}}}/
                  {{\nu}_{\rm{v}}}]^{{-\alpha}(\rm{radio},\rm{v})}
   \end{equation}
    $S_{\rm{radio}}$ represents the modeled flux density at the radio 
    frequencies
    of 8, 14.5, 22 and 37\,GHz. $S_{\rm{v}}$ represents the modeled 
    V-band optical flux density. The modeled spectral indexes
    $\alpha$(\rm{radio},\rm{v}) are selected
    to reasonably fit the observed radio/mm light-curves (see Tables 5 
   and 6). This method is equivalent to select appropriate two-point
   radio spectral indexes ${\alpha}$(22GHz,37GHz), ${\alpha}$(14.5GHz,37GHz)
    and $\alpha$(8GHz,37GHz) to simulate the radio light-curves 
   at 22, 14.5 and 8\,GHz, when the 37\,GHz light-curve is fitted by using
   $\alpha$(37GHz,V-band).
   \subsection{Cosmological model}
    In this work, we adopt a $\Lambda$CDM cosmological model with the
   parameters as: ${\Omega}_m$\,=\,0.27, ${\Omega}_{\Lambda}$\,=\,0.73 and
   $\rm{H_0}$=71\,km\,${\rm{s}}^{-1}$\,${\rm{Mpc}}^{-1}$ (Spergel et al.
    \cite{Sp03}, Komatsu et
    al. \cite{Ko09}). 1\,mas\,=\,4.5\,pc (Hogg, \cite{Ho99}).
   \begin{figure*}
   \centering
   \includegraphics[width=8cm,angle=-90]{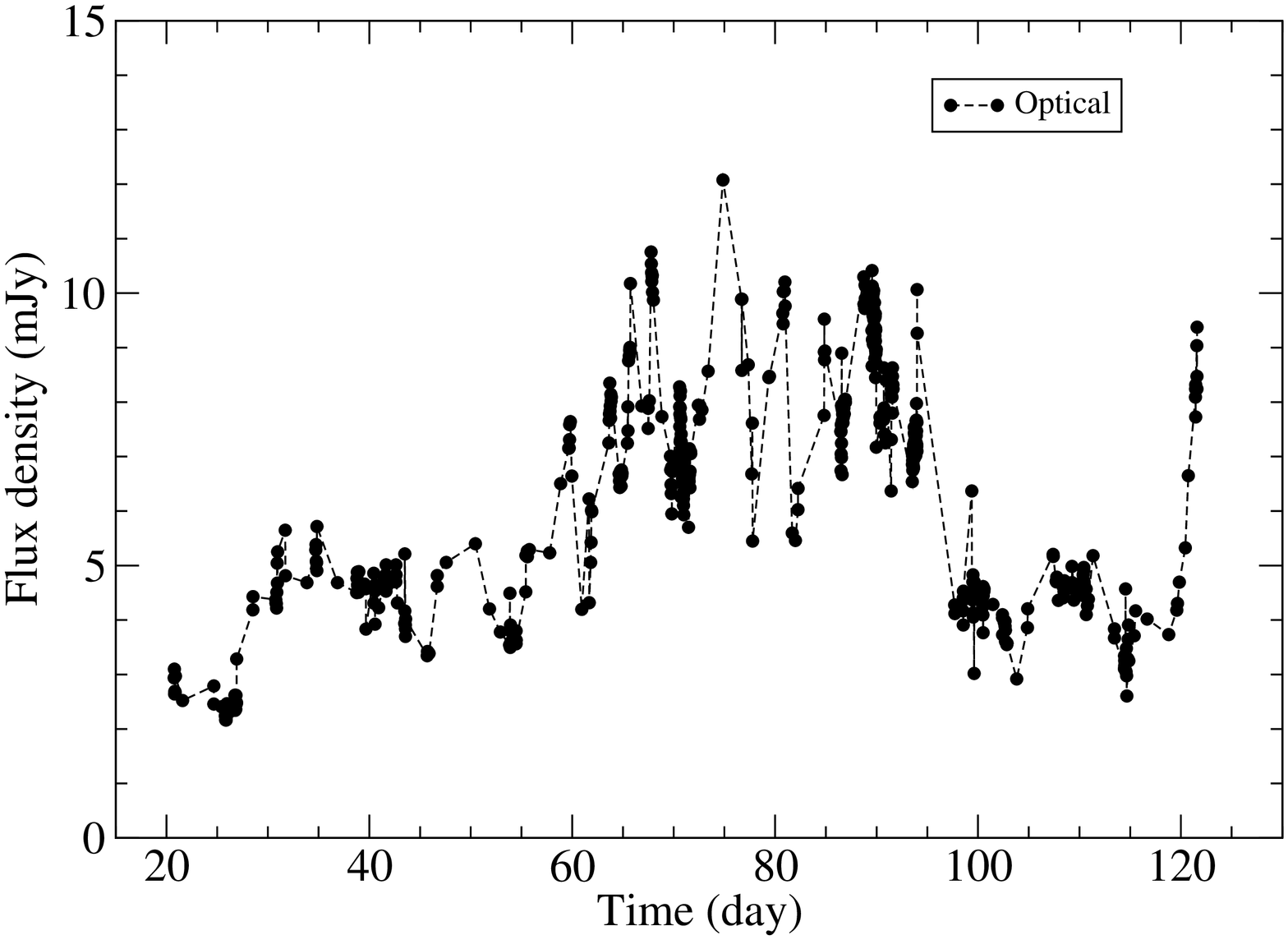}
   \includegraphics[width=8cm,angle=-90]{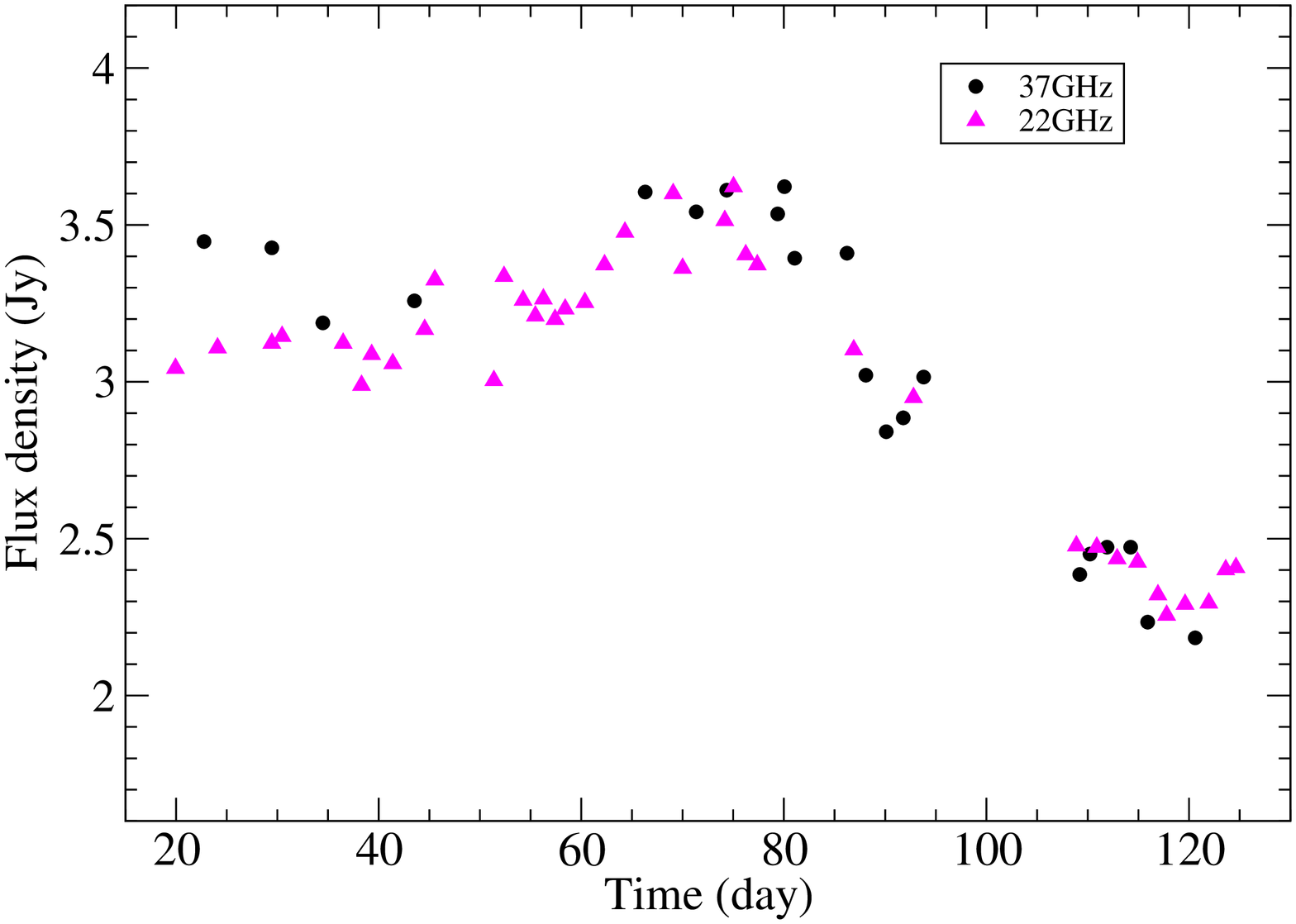}
   \includegraphics[width=8cm,angle=-90]{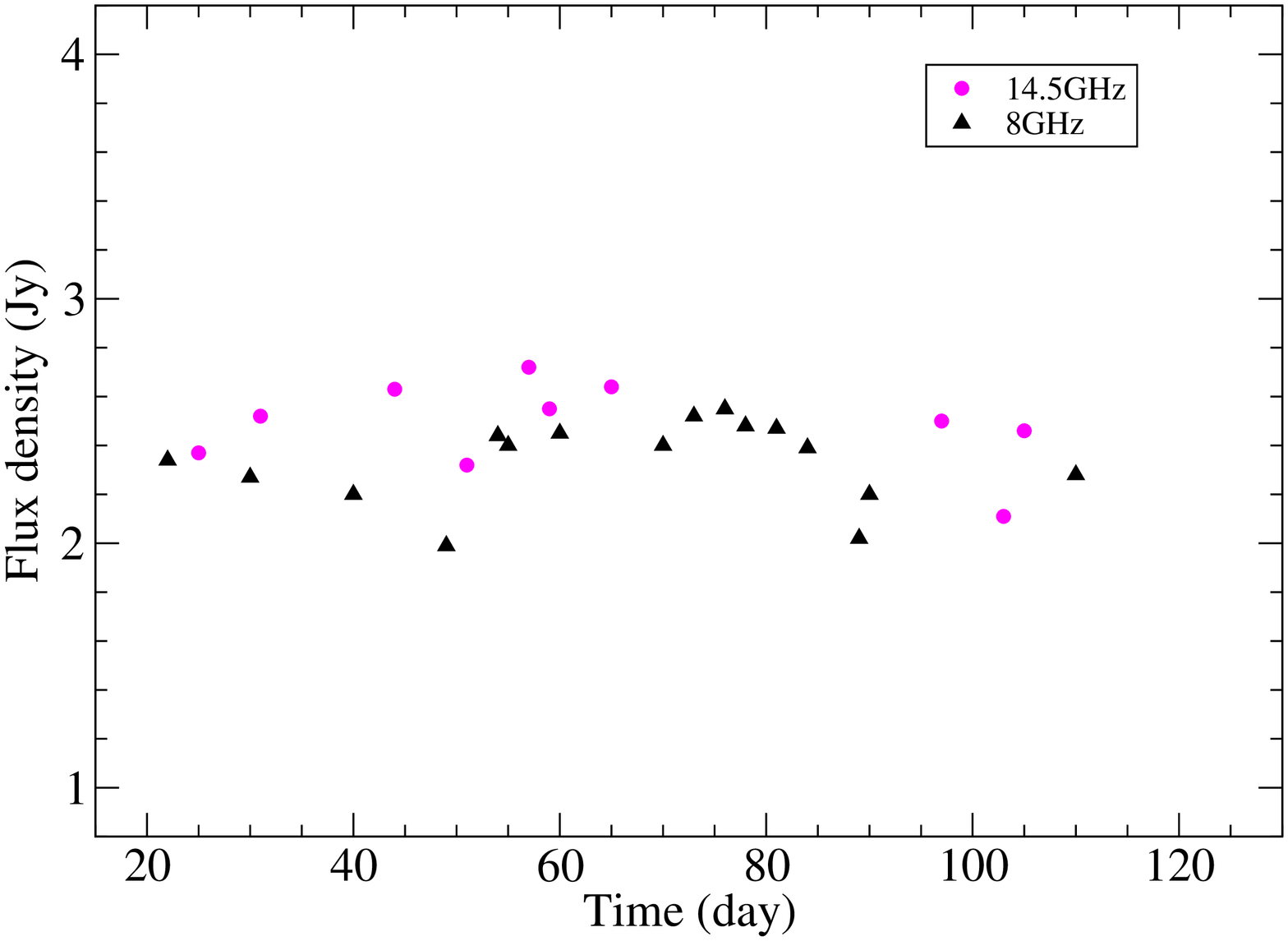}
   \caption{Quasi-periodic optical outburst in 1995.84: 
   optical V-band light curve 
  (upper panel) and radio/millimeter light curves observed at 37/22\,GHz
   (middle panel) and 14.5/8\,GHz (bottom panel). Zero point of  
   time=Julian date 2450000 (1995.7735).}
   \end{figure*}
\section{Connection between optical and radio/millimeter outbursts}
   The light curves observed during the period 1995.77 (JD2450000)
   -- 1996.13 (JD2450130) in the optical V-band and radio/mm bands (37, 22,
   14.5 and 8\,GHz) are shown in Figure 3. The V-band data were taken from
  Basta \& Hudec (\cite{Ba06}) and OJ-94 project (Nilsson, 
   private communication). The 37 and 22\,GHz data were taken from Valtaoja 
  et al. (\cite{Val20}) and the 14.5 and 8\,GHz data from UMRO-archive
   (Aller, private communication).\\
   It can be seen that the optical light curve consists of a large number
   of elementary spike-like flares with time scales of days. 
    Figure 3 shows that the envelopes of the radio/mm light curves  are
   much more smoother than that of the optical light-curve. This is due to 
   the local spectral indexes at radio/mm wavelengths are much smaller than
   that at optical V-band (${\alpha}_{\rm{v}}$=1.0). However, even by 
   visual inspection it can be seen that the radio/mm light-curves have their
    envelopes very similar to that of the optical envelope during the period 
    JD2450060--2450090. Very clearly, for the 37 and 22\,GHz light curves 
    this similarity in the envelopes can be extended to JD2450120 when both
    the optical flux density and the 37--22\,GHz flux densities 
    simultaneously reached to their
     quiet levels. Moreover, a few optical spikes (e.g, at JD2450075 and
     JD2450080) can be recognized to have their concurrent flares in
     the 37 and 22\,GHz light-curves. All these similarities in the 
     radio/mm  and the optical variations clearly demonstrate that the 
    radio/mm flares are simultaneous with the optical flares without 
    measurable time delays (less than two days), as suggested previously
    by Valtaoja et al. (\cite{Val20}).\\
    The close connection 
   between the radio/mm outbursts at four frequencies (37, 22, 14.5 and 8\,GHz)
   and the optical outburst (or simultaneous variability behavior) during 
   the period of 1995.8--1996.1 provides further evidence for 
   the nonthermal nature of the quasi-periodic optical outbursts in OJ287, 
   in addition to the association of the high energy $\gamma$-ray and optical
    outbursts.
   \footnote{Also see the results of the
   correlated optical and multi-waveband radio/mm outbursts observed in
   1971--1974 (Usher \cite{Us79}).} The close connection between the
   optical and radio/mm outbursts may be regarded as another decisive 
   evidence that the optical outbursts and their constitutive (elementary)
   flares in OJ287 are synchrotron in origin. The arguments against lighthouse
   models (e.g., Valtaoja et al. \cite{Val20}) can now be clarified.\\
    (1) Firstly, in the precessing jet nozzle scenario, both the optical and 
    radio/mm outbursts are due to enhanced Doppler boosting induced by  
    lighthouse effect due to the helical motion of the same superluminal knots.
    Thus the radio emissions are similarly boosted.  Due to the nature of
    synchrotron emission the polarization properties in optical and radio/mm
    wavebands can be well interpreted.\\
    (2) The interpretation of both the optical and radio/mm outbursts in terms
     of the precessing jet nozzle models does not involve precession of the 
    whole jet. Thus the emissions in optical and radio/mm wavebands
     from the underlying jet (or the quiescent base-levels) do not experience
     Doppler boosting during the outbursts.\\
    (3) The lighthouse model proposed here requires the emitting source
     having some kind of radiation structure (or distribution):  the radio/mm 
    and optical emission do not come from the exactly same region. Most
    possibly, it could be a planar shock front with a core-envelope structure. 
    Its central (core) region dominates the emission in IR-optical bands (and
    the associated $\gamma$-rays), while its outer region (envelope)
    dominates the
     emission in radio/mm wavebands. Thus the optical  and radio/mm emissions 
     can be simultaneously and similarly Doppler boosted, when the source moves
     along helical trajectories, if the shock front is perpendicular to the
     direction of the motion. This core-envelope structure is consistent with
     the concept of the stratification of magnetic surface structure in the
     magnetospheres of spinning black hole-accretion disk systems
     (e.g., Camenzind,\cite{Cam90}).\\
    (4) The radio/mm emitting regions may have different optical depths or
     opacity effects in  different cases. This can explain why some 
     quasi-periodic optical outbursts do not have radio counterparts.
     But even in these cases delayed radio outbursts  are usually 
     observed if the optical knots evolve to form pure radio knots in
     outer jet regions on parsec scales (Valtaoja et al.
     \cite{Val20}, Tateyama et al. \cite{Ta99}, Britzen et al. \cite{Br18},
     Qian \cite{QiXiv18}).\\
     (5) Our studies (Qian \cite{QiXiv19} and this work) have found that 
    the quasi-periodic optical outbursts can be decomposed into several
     elementary flares, each having symmetric profiles with similar rising 
    and declining timescales. This explains why the individual flares  during 
    quiescent period and flaring period have similar variability
    timescales (Valtaoja et al. \cite{Val20}), because the elementary flares
     during quiescent and flaring periods have similar properties: 
    flux density is boosted and timescale is compressed by Doppler effects.\\
     (6) Since the flux enhancements induced by Doppler boosting depend on
     the local spectral index at the emission frequency, 
     the flattening of the radio/mm spectrum with respect to the IR-optical
      spectrum results in the amplitude of the radio/mm outbursts being much
      smaller than that of the optical outbursts, as usually observed.\\
    (7) This work only investigate the interpretation of the optical and 
     radio/mm outbursts (related to superluminal knots) on timescales of
     $\sim$10--100\,days. Thus the underlying jet flux density 
    (or the base-level 
     flux density) is assumed to be constant. However, the physical conditions
     of the jet (jet bulk motion and magnetic field structure, etc.) should
     vary on longer time-scales and the jet precession (with a period 
     of $\sim$12\,yr) will also modulate the 
     underlying jet emission. In order to understand the whole phenomena in
     OJ287 all these ingredients should be taken into consideration.\\
    In the following we will model simulate the optical and radio/mm light
    curves  in terms of lighthouse effect due to the helical motion of the
    superluminal optical/radio knots, respectively.
     \begin{table*}
   \caption{Model-simulation parameters for the 36 elementary flares at V-band.
    $t_{\rm{peak}}$ (JD--2450000) -- peaking epoch of the modeled 
    elementary flares: 1995.840 is
    corresponding to Julian date (JD) 2450024.29. Base-level flux 
    density $S_{\rm{v,b}}$=2.70\,mJy. $S_{\rm{v,p}}$ (mJy) -- modeled 
    peak flux density.  $\Gamma$ -- bulk Lorentz factor of the optical knot.
     ${\delta}_{\rm{max}}$ -- maximum Doppler
    factor, r=${\delta}_{\rm{max}}$/${\delta}_{\rm{min}}$, 
    $S_{\rm{v,co}}$ (mJy) -- modeled 
    comoving (or deboosted) flux density. FWHM (day) -- full width at 
     half-maximum of the elementary flare profile.}
   \begin{flushleft}
   \centering
   \begin{tabular}{lllllll}
   \hline
   $t_{\rm{peak}}$ & $S_{\rm{v,p}}$(mJy) &  $\Gamma$ & ${\delta}_{\rm{max}}$ &
          r & $S_{\rm{v,co}}$(mJy) & FWHM \\
   \hline
    27.86 & 6.54 &  17.0 & 33.58 & 10.95 & 3.03(-6) & 1.0 \\
    31.36 & 6.54 &  17.0 & 33.54 & 10.94 & 3.03(-6) & 1.0 \\
    34.46 & 6.54 &  17.0 & 33.54 & 10.94 & 3.03(-6) & 1.0 \\
    37.66 & 6.54 &  17.0 & 33.54 & 10.94 & 3.03(-6) & 1.0 \\
    39.26 & 6.54 &  17.0 & 33.54 & 10.94 & 3.03(-6) & 1.0 \\
    41.06 & 6.54 &  17.0 & 33.54 & 10.94 & 3.03(-6) & 1.0 \\
    42.86 & 6.54 &  17.0 & 33.54 & 10.94 & 3.03(-6) & 1.0 \\
    44.56 & 6.54 &  17.0 & 33.54 & 10.94 & 3.03(-6) & 1.0 \\
    46.41 & 6.54 &  17.0 & 33.54 & 10.94 & 3.03(-6) & 1.0 \\
    47.76 & 6.54 &  17.0 & 33.54 & 10.94 & 3.03(-6) & 1.0 \\
    50.81 & 6.54 &  17.0 & 33.54 & 10.94 & 3.03(-6) & 1.0 \\
    52.21 & 6.54 &  17.0 & 33.54 & 10.94 & 3.03(-6) & 1.0 \\
    54.76 & 6.54 &  17.0 & 33.54 & 10.94 & 3.03(-6) & 1.0 \\
    56.35 & 6.43 &  14.0 & 27.73 & 7.76  & 6.32(-6) & 1.8 \\
    59.29 & 7.47 &  12.5 & 24.79 & 6.39 & 1.26(-5) & 2.5 \\ 
    63.07 & 7.83 &  13.0 & 25.77 & 6.83 & 1.16(-5) & 2.3 \\
    65.95 & 9.04 &  14.0 & 27.73 & 7.76 & 1.07(-5) & 1.8 \\
    68.12 & 12.0 &  14.5 & 28.70 & 8.25 & 1.37(-5) & 1.5 \\
    70.30 & 10.9 &  17.5 & 34.51 & 11.5 & 5.81(-6) & 0.9 \\
    71.95 & 10.4 &  16.5 & 32.58 & 10.4 & 6.82(-6) & 1.0 \\
    75.05 & 12.3 &  11.5 & 22.82 & 5.56 & 3.53(-5) & 3.3 \\
    77.06 & 6.61 &  15.0 & 29.67 & 8.75 & 5.05(-6) & 1.3 \\
    80.32 & 11.7 &  14.5 & 28.70 & 8.25 & 1.32(-5) & 1.6 \\
    85.16 & 8.79 &  12.0 & 23.81 & 5.97 & 1.89(-5) & 3.0 \\
    88.96 & 10.1 &  12.0 & 23.81 & 5.97 & 2.32(-5) & 3.0 \\
    92.12 & 9.22 &  14.5 & 28.70 & 8.25 & 6.66(-6) & 1.6 \\
    94.98 & 10.4 &  13.5 & 26.75 & 7.29 & 1.50(-5) & 2.2 \\
    99.06 & 6.54 &  17.0 & 33.54 & 10.94 & 3.03(-6) & 1.0 \\
    101.02 & 6.54 & 17.0 & 33.54 & 10.94 & 3.03(-6) & 1.0 \\
    103.17 & 6.54 & 17.0 & 33.54 & 10.94 & 3.03(-6) & 1.0 \\
    105.62 & 6.54 & 17.0 & 33.54 & 10.94 & 3.03(-6) & 1.0 \\
    107.11 & 6.54 & 17.0 & 33.54 & 10.94 & 3.03(-6) & 1.0 \\
    108.90 & 6.54 & 17.0 & 33.54 & 10.94 & 3.03(-6) & 1.0 \\
    111.15 & 6.54 & 17.0 & 33.54 & 10.94 & 3.03(-6) & 1.0 \\
    113.83 & 6.54 & 17.0 & 33.54 & 10.94 & 3.03(-6) & 1.0 \\
    116.17 & 6.54 & 17.0 & 33.54 & 10.94 & 3.03(-6) & 1.0 \\
    122.10 & 10.4 & 13.5 & 26.75 & 7.29 & 1.50(-5) & 2.2 \\
   \hline
   \end{tabular}
   \end{flushleft}
   \end{table*}
 \subsection{Model simulation of the optical light-curve}
    Since the quasi-periodic optical outbursts can be decomposed into several
    elementary outbursts and each of the elementary flares can be interpreted
    in terms of the helical motion model, we can use  the processing jet nozzle
    model and the kinematic properties observed on parsec scales to investigate
    the light-curves of optical outbursts in OJ287.\\
     The model simulation of the optical light-curve for the 1995.84 outburst
     event is shown in Figure 4.  The model parameters and results for all 
    the optical elementary flares are summarized in Table 5: 
      $t_{\rm{peak}}$ (JD--2450000) -- peaking epoch of elementary flare;
      $S_{\rm{v}}$ (mJy) -- optical flux density at peak;
      $S_{\rm{b}}$ (mJy) -- base-level (underlying jet) flux density;
      $\Gamma$ -- bulk Lorentz factor of superluminal optical knots;
      ${\delta}_{\rm{max}}$ -- maximum Doppler factor;
      $\rm{r}$ -- ratio of the maximum and minimum Doppler factor;
      $S_{\rm{v,co}}$ (mJy) -- comoving (or de-boosted) flux density and the 
       numbers in the parentheses represent the powers of 10;
      FWHM (day) -- full width at half-maximum of flare profile.\\
      Due to the insufficient sampling of the optical observations it is very
     difficult to recognize the discrete elementary flares. The
      simulation is certainly not unique, depending on personal assessments on
     the decomposition. We have tried to recognize the symmetric
     distributions of the data-points. The simulation results described here 
     may provide us
      useful information about the elementary flares: timescale, bulk motion 
     Lorentz factor,  Doppler beaming factor,
     comoving optical flux density etc., which may be useful to understand
     the nature of the optical outbursts, jet physics, 
     superlumianl knot kinematics and the characteristics of
     the putative central black hole binary system.\\
     For describing the simulation results we divide the light-curve
     into three segments: segment I for time-interval JD2450024--2450049
     (1995.8392--1995.9077; 25\,days), 
     segment II for time-interval JD2450050--2450097 
     (1995.9104--1996.0391; 47\,days) and segment III for 
    time-interval JD2450098--2450118 (1996.0418--1996.0966; 20\,days).
    \subsubsection{Segment I}
     This part of the light-curve may be regarded as the starting 
    stage of the 1995.84 outburst event. Epoch JD2450024 (1995.841)
     has been predicted as the starting epoch of the primary thermal 
     outburst in the disk-impact model, but this impact-outburst 
    did not happen and the source
    retained at a low activity level within the first 25\,days: neither
    a strong burst nor a standard impact-burst light curve was observed, while
    the follow-up synchrotron flares happened as usual.
    The disappearance of the primary impact-outburst seems difficult to be
    understood under the disk-impact scenario, and the claimed timing of this 
   impact-outburst seems losing its physical meaning.\\
      Applying the lighthouse model we use ten elementary flares with 
    similar symmetric profiles to  fit the light-curve, which is
     displayed in upper panel of Figure 4.
     Although the data sampling is very sparse, the elementary flares 
    (at $t_{\rm{peak}}$=27.86, 31.36, 34.46, 39.26, 41.06, 42.86 and 46.41)
     can reasonably well fit the data-points, implying 
     steady ejections of superluminal optical knots with similar properties 
     during this time-interval. All the optical knots are modeled with the same 
      bulk Lorentz factor of $\Gamma$=17, having time-scales of $\sim$1\,day.
    \subsubsection{Segment II}
     This segment of the light-curve may be regarded as the main part of the 
     optical event starting in 1995.84, revealing a group of large
     amplitude spike-like flares and an envelope of typical pattern of
     rising-plateau-declining stages. Although  the data sampling
     is still insufficient to accurately decide the peak epoch and amplitude
     of these spikes, the decomposition of the outburst into its elementary
     flares can be made appropriately. Seventeen elementary flares with 
     different Lorentz factors (in the range of 11.5 to 17.5) have been used
     (see Table 5). The simulation results are displayed
      in the middle panel of Figure 4. The bold
     violet curve denotes the modeled total flux density, which reasonably 
     well fit most of the observational data-points. The elementary flares
     have their time-scales in the range of 0.9 to  3.3\,days.
    \subsubsection{Segment III}
     This segment may be regarded as the ending stage of the outburst
     event, when the flaring activity sharply declined. Nine elementary
     flares with similar symmetric profiles 
     and Lorentz factors ($\Gamma$=17) has been used to simulate
     the light-curve. Although the data sampling is insufficient to 
     accurately decide their peak amplitudes,
     the elementary flares (at $t_{\rm{peak}}$=99.06,
     101.02, 108.90, 111.15, 113.83 and 116.17) can reasonably well fit the
     observational data-points.  After JD2450120 
    another activity period started.\\
    As a brief summary, it can be      seen that the entire light-curve of the
    optical outburst event staring in 1995.84 (JD2450024) during a period of
    $\sim$3 months can be model-simulated in terms of our
    lighthouse model (helical motion model). All these elementary flares
    have symmetric profiles with similar Lorentz factors and time-scales.
    Their comoving (or de-boosted) flux densities
    are on order of $10^{-6}$\,mJy, implying that the Doppler boosting 
    effects are very strong. The characteristics of the superluminal 
    optical/radio knots and their helical motion explored in this work 
    may be significant for understanding the nature of the whole phenomena 
    in OJ287.
    \begin{figure*}
    \centering
    \includegraphics[width=8cm,angle=-90]{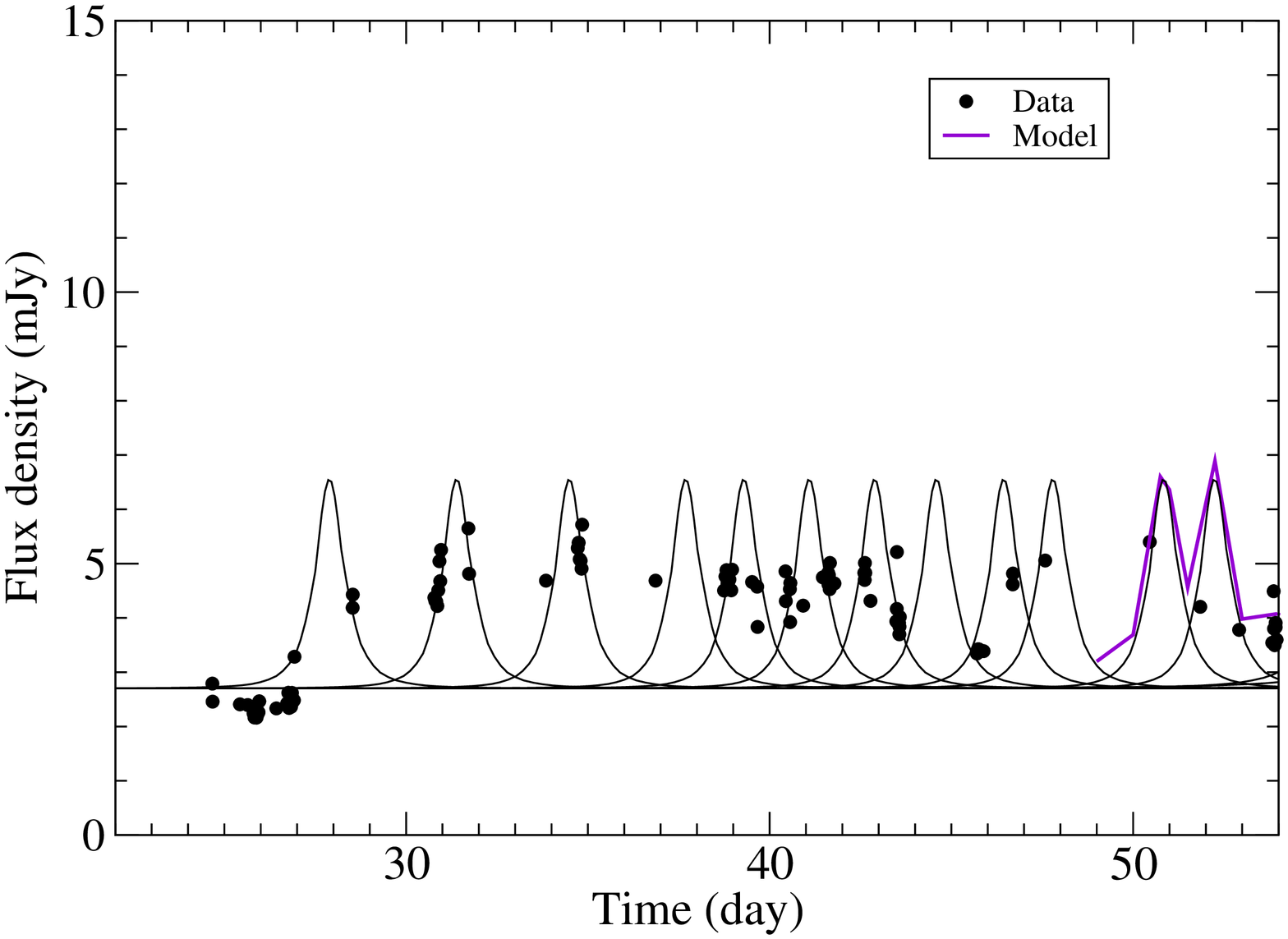}
    \includegraphics[width=8cm,angle=-90]{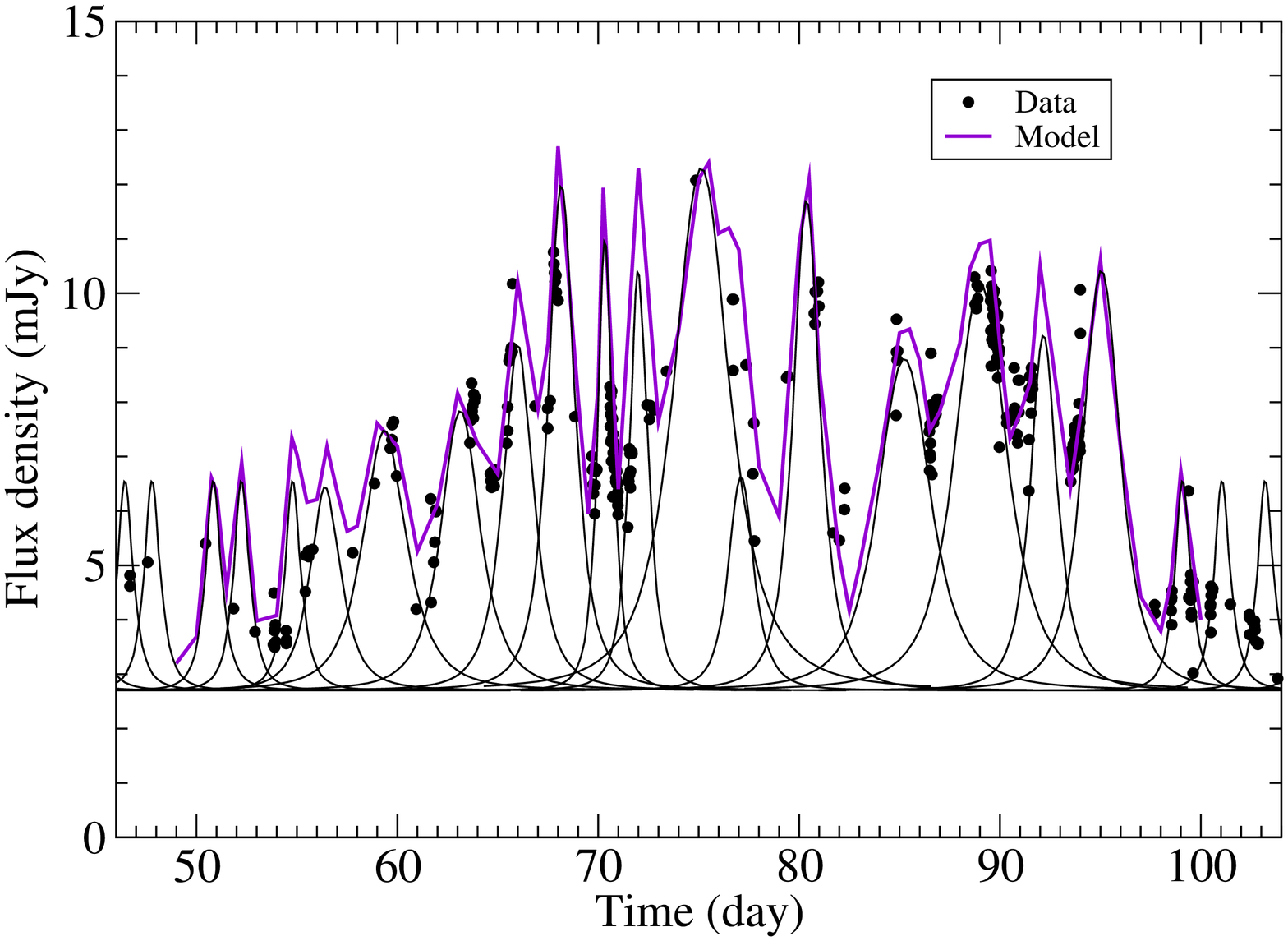}
    \includegraphics[width=8cm,angle=-90]{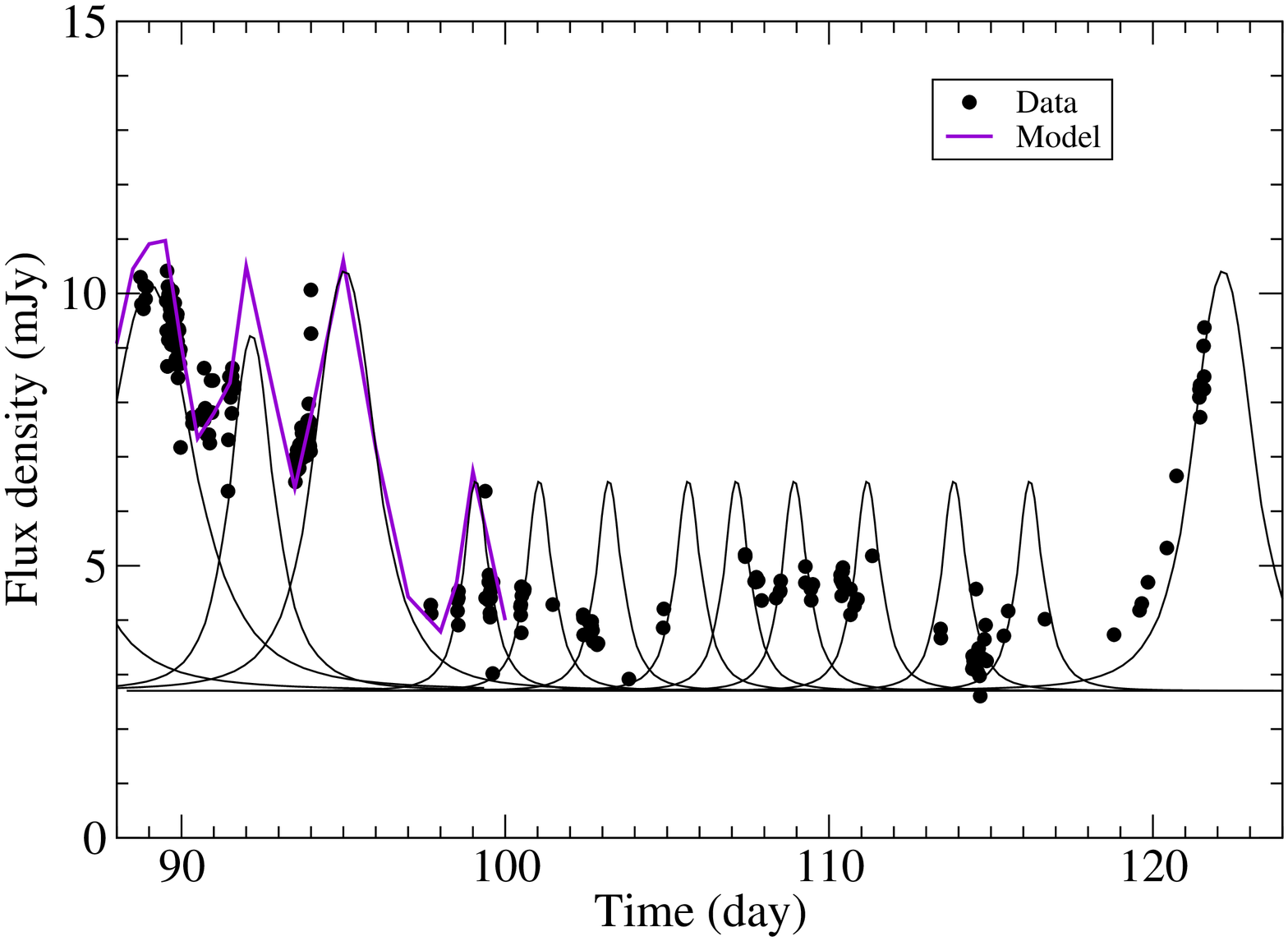}
    \caption{Model simulation of the optical light curve for the 1995.8 
    optical outburst event (1995.84--1996.10): the light curve is
    displayed in three segments, separately. The model parameters are
    listed in Table 5.}
    \end{figure*}
  \subsection{Model simulation of 37\,GHz light curve}
    Corresponding to the simulation of the optical light-curve,
    the 37\,GHz light-curve has been simulated and shown in Figure 5.
    The modeled 37\,GHz light curve 
     is derived from the modeled optical light-curve by selecting the
    spectral index $\alpha$(37GHz,V-band) between 37\,GHz and optical
     V-band for each of the elementary flares which are listed in Table 6.
    The 22\,GHz light-curve can be fitted using a similar method by selecting
    spectral index $\alpha$(22GHz,V-band) for the elementary flares, but
    the fitting results are not given here. However, the 22\,GHz data-points
    are also shown in Figure 5 to help visual-inspecting the quality of the
    simulation of the 37\,GHz light-curve. The simulation results are described
    separately for three segments as follows.   
  \subsubsection{Segment I}
    Corresponding to the same segment of the optical light-curve 
    (JD2450024--2450049), ten elementary flares at 37\,GHz are simulated
    (upper panel of Figure 5). Unfortunately, only three data-points are
     available, but the data-point at $t_{\rm{peak}}$=34.46
     is well fitted, coincided with the optical flare. There are 
    four 22\,GHz data-points very close to the profile of the flares at 
    $t_{\rm{peak}}$=37.66, 39.26, 41.06 and 44.56, indicating that the optical
     flares should have their counterparts at 37\,GHz. 
     The 37 and 22\,GHz data-points between 2450028 and 2450030
     have no optical counterpart, possibly implying the existence of 
     a pure radio/mm burst. 
  \subsubsection{Segment II}
   Corresponding to the main part of the optical light-curve seventeen
   37\,GHz elementary flares are simulated and the results are shown in
   the middle panel of Figure 5. The bold violet curve denotes the modeled
   total flux density curve. It can be seen that almost all the 37\,GHz 
   data-points are reasonably well fitted (within uncertainty of 
   $\sim$1-2 days), indicating their simultaneous flaring at 
    optical and 37\,GHz wavelengths.
   In particular, the optical peak at $t_{\rm{peak}}$=80.32 has a well 
   fitted 37\,GHz peak. During the time-interval between 2450050 and 2450078, 
    the 22\,GHz data-points are close to the profiles of the 37\,GHz elementary
    flares, demonstrating the  simultaneous variations at 37/22\,GHz 
    and optical wavelengths.
    \subsubsection{Segment III}
    Corresponding to the same part of the optical light-curve, nine 37\,GHz
    elementary flares have been used in the simulation. The results
     are displayed in 
    the bottom panel of Figure 5. Only five data-points are available
     and four elementary flares are invoked, having very low
    flux densities and a spectral index ${\alpha}$(37GHz,V-band) of 
    0.40, much smaller than those (0.50--0.58) during the outburst peaking 
    stage (between 2450050 to 2450085; see Table 6). 
    It can be seen from Figure 5 (middle panel) that the observed 22\,GHz 
    light-curve is slightly below the 37\,GHz light-curve, thus introduction 
    of  spectral index $\alpha$(22GHz,V-band) slightly smaller than 
    the $\alpha$(37GHz,V-band), will well fit
    the 22\,GHz light curve.\\
     As shown above, we can  reasonably well explain the optical 
    and 37--22\,GHz 
    light curves in terms of the lighthouse model under the precessing 
    jet nozzle scenario.
     Each of the optical elementary flares have their 37/22\,GHz counterparts.
     Same Doppler boosting factors are effective for both optical and radio/mm
     flares. These results can not be obtained in the whole-jet
     precessing models (e.g., Valtaoja et al. \cite{Val20}).\\
     The significant point is that the 37--22\,GHz elementary flares are
     shown to be synchronous with the optical ones. If the 37--22\,GHz flares 
     were delayed with respect to the
     optical flares, the spikes on time-scales of $\sim$1-2\,days
     in the 37--22\,GHz 
     light-curves might have been smoothed out and have not been observed.\\
     \begin{table*}
    \caption{Model-simulation parameters for the thirty-six 
     elementary flares at 37\,GHz. $S_{\rm{b,37}}$=2.31\,Jy.}
    \begin{flushleft}
    \centering
    \begin{tabular}{llll}
    \hline
    $t_{\rm{peak}}$ & $S_{37,\rm{p}}$ & ${\alpha}_{37,\rm{v}}$ & 
                      $S_{37,\rm{co}}$ \\
    \hline
    27.86 & 3.33 &  0.58 & 8.00(-7) \\
    31.36 & 3.33 &  0.58  & 8.00(-7)  \\
    34.46 & 3.33 &  0.58  & 8.00(-7)  \\
    37.66 & 3.33 &  0.58 & 8.00(-7) \\
    39.26 & 3.33 &  0.58  & 8.00(-7) \\
    41.06 & 3.33 &  0.58  & 8.00(-7) \\
    42.86 & 3.33 &  0.58  & 8.00(-7) \\
    44.56 & 3.33 &  0.58  & 8.00(-7)  \\
    46.41 & 3.33 &  0.58  & 8.00(-7)  \\
    47.76 & 3.33 &  0.58  & 8.00(-7)  \\
    50.81 & 3.33 &  0.58  & 8.00(-7) \\
    52.21 & 3.33 &  0.58  & 8.00(-7) \\
    54.76 & 3.33 &  0.58 & 8.00(-7) \\
    56.35 & 3.30 &  0.58 & 1.67(-6)  \\
    59.29 & 3.35 &  0.56  & 2.76(-6)  \\
    63.07 & 3.43 &  0.56  & 2.54(-6)  \\
    65.95 & 3.45 &  0.54 & 1.93(-6) \\
    68.12 & 3.45 &  0.50  & 1.69(-6) \\
    70.30 & 3.42 &  0.51  & 7.85(-7)  \\
    71.95 & 3.35 &  0.51   & 9.22(-7) \\
    75.05 & 3.61 &  0.51  &  4.77(-6)  \\
    77.06 & 2.84 &  0.51  & 6.81(-7) \\
    80.32 & 3.52 &  0.51  &  1.79(-6) \\
    85.16 & 3.41 &  0.54  &  3.42(-6) \\
    88.96 & 2.82 &  0.44  &  1.59(-6) \\
    92.12 & 2.80 &  0.45  &  7.22(-7)  \\
    94.98 & 3.35 &  0.51  &  2.03(-6)  \\
    99.06 & 2.83 &  0.51 & 4.11(-7) \\
    101.02 & 2.83 & 0.51 & 4.11(-7)  \\
    103.17 & 2.83 & 0.51 &  4.11(-7) \\
    105.62 & 2.83 & 0.51 & 4.11(-7) \\
    107.11 & 2.83 & 0.51 & 4.11(-7) \\
    108.90 & 2.49 & 0.40 & 1.42(-7) \\
    111.15 & 2.49 & 0.40 & 1.42(-7)  \\
    113.83 & 2.49 & 0.40 & 1.42(-7) \\
    116.17 & 2.49 & 0.40 & 1.42(-7) \\
    122.10 & 2.36 & 0.20 & 0.98(-7) \\
    \hline
    \end{tabular}
    \end{flushleft}
    \end{table*}
    \begin{figure*}
    \centering
    \includegraphics[width=8cm,angle=-90]{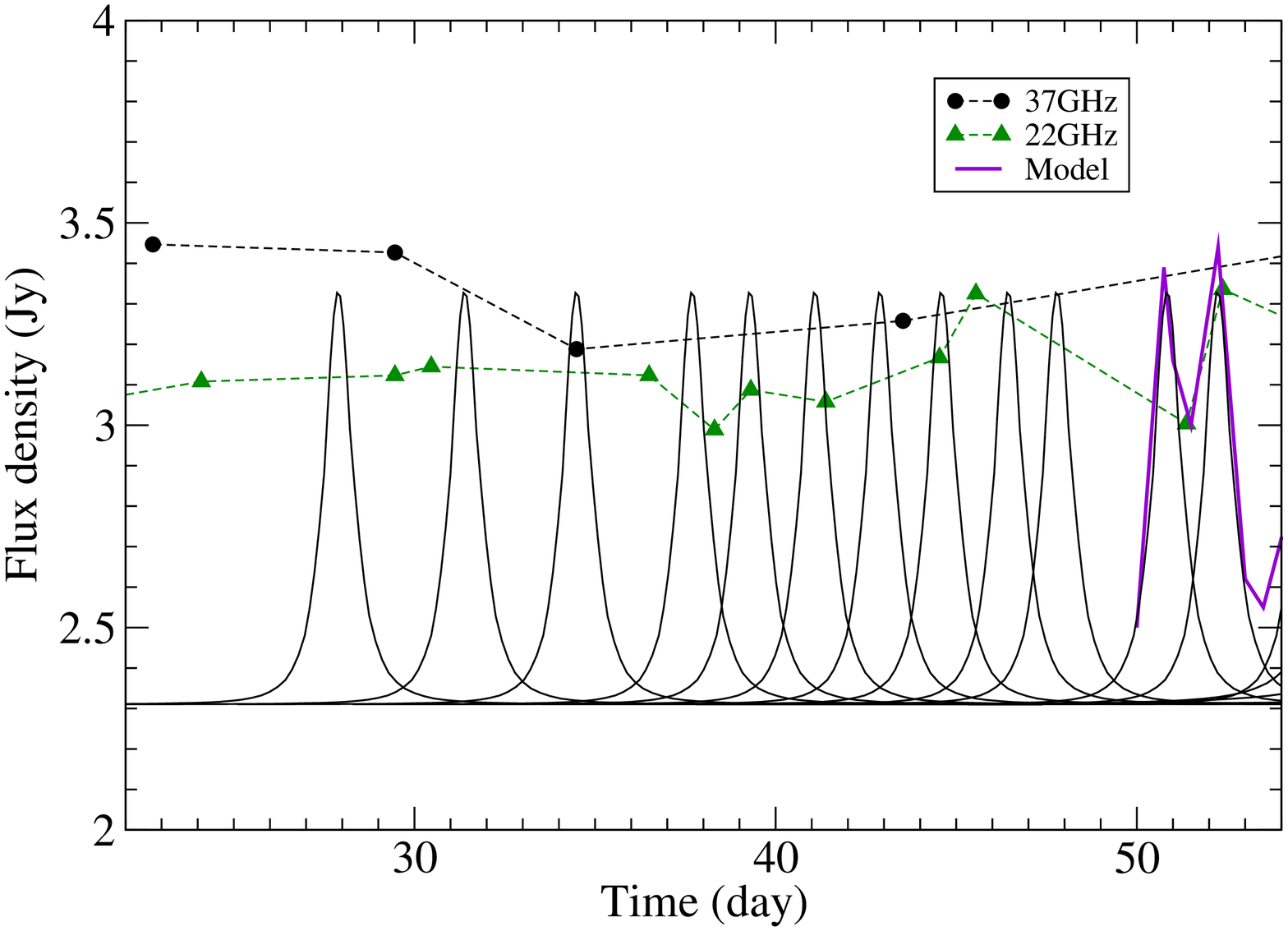}
    \includegraphics[width=8cm,angle=-90]{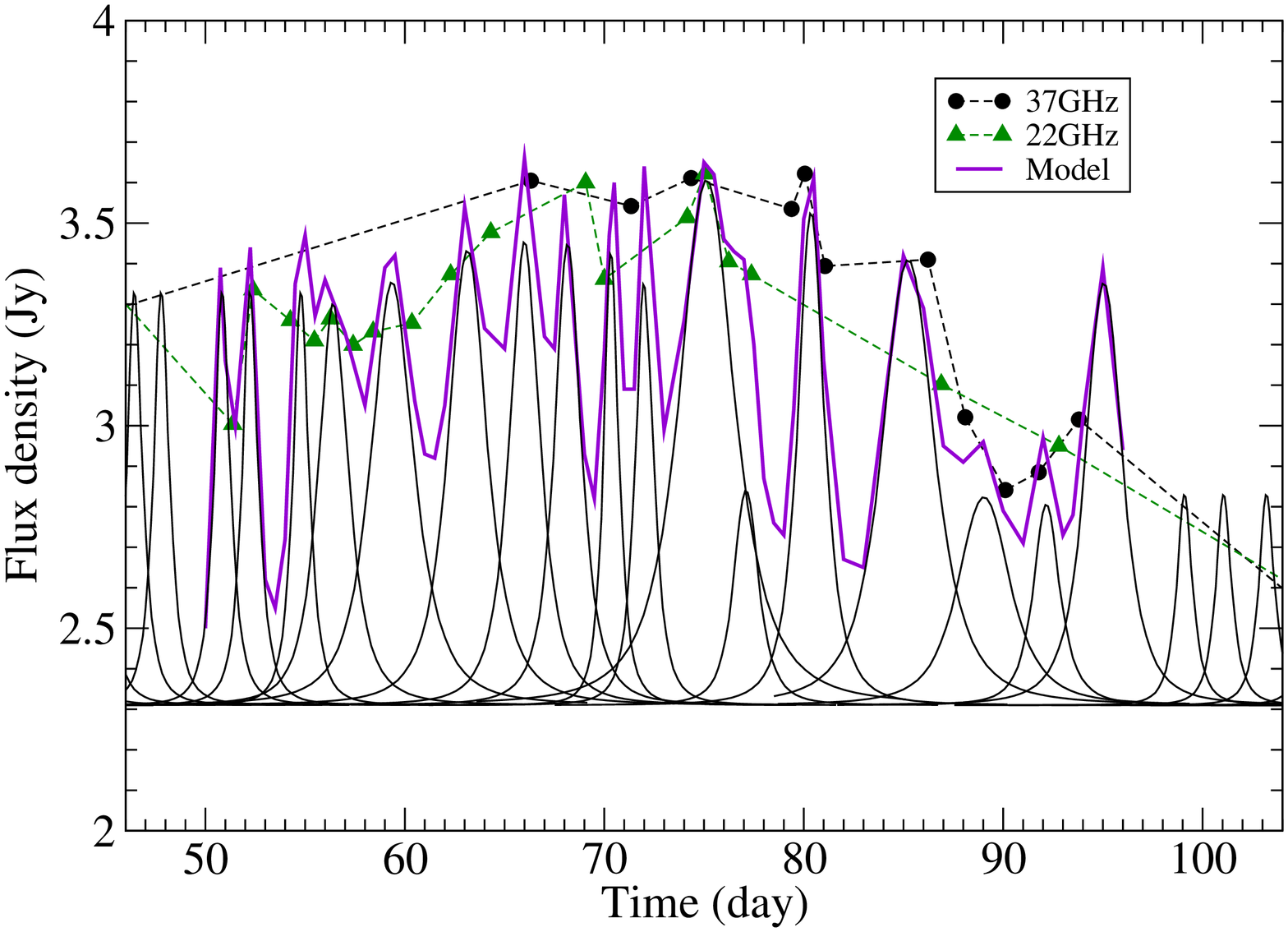}
    \includegraphics[width=8cm,angle=-90]{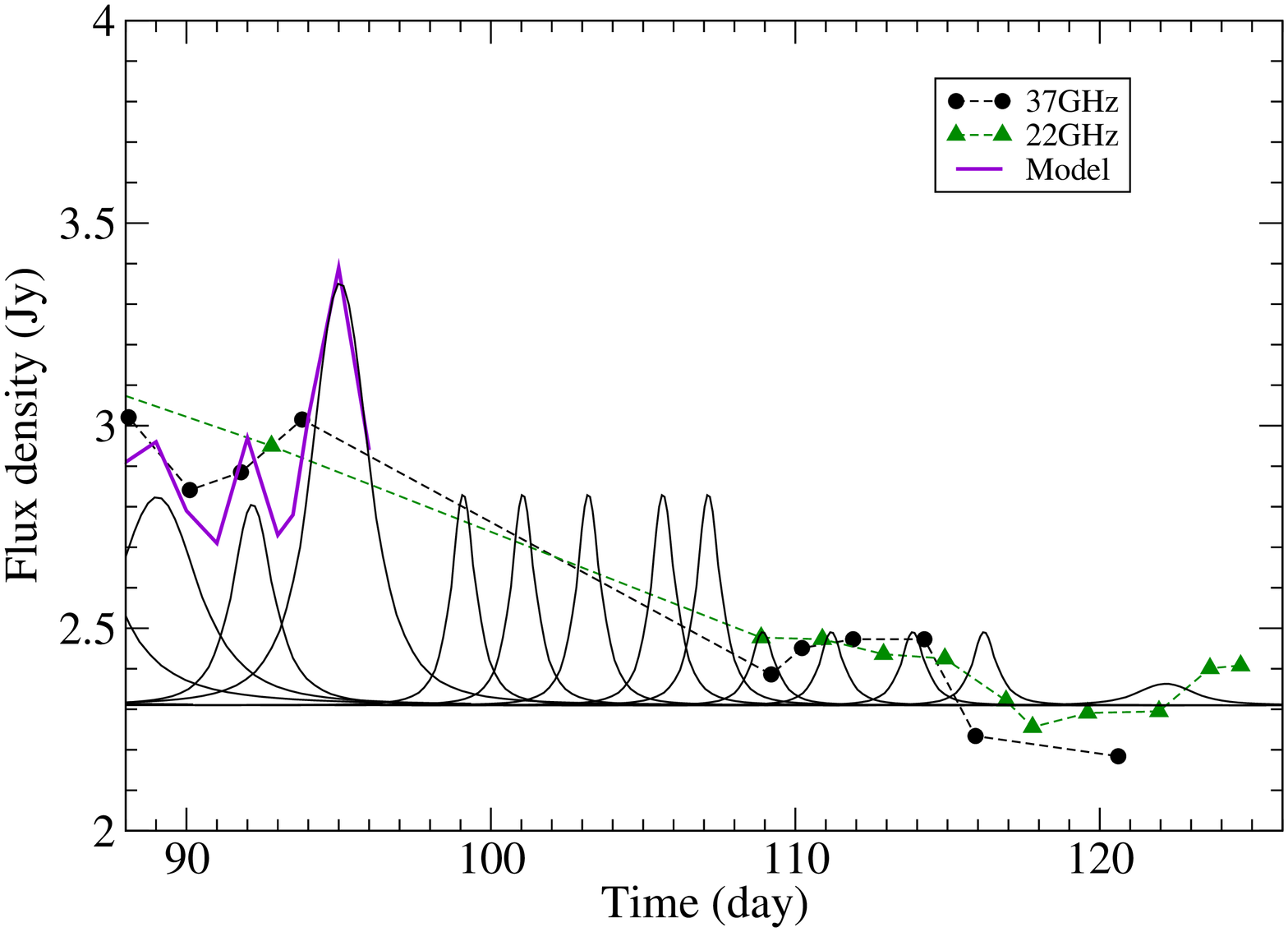}
    \caption{Model simulation of the 37\,GHz light curve during the 
    outburst event between  1995.84 and 1996.10 (JD2450024--2540118).
    The model parameters are listed in Table 6.}
    \end{figure*}
     \begin{table*}
    \caption{Model-simulation parameters for the thirty-six elementary 
     flares at 14.5\,GHz. $S_b$=1.30\,Jy.}
    \begin{flushleft}
    \centering
    \begin{tabular}{llll}
    \hline
    $t_{\rm{peak}}$ & $S_{15,\rm{p}}$ & ${\alpha}_{15,\rm{v}}$ & 
                $S_{\rm{co},15}$  \\
    \hline
    27.86 & 2.45  & 0.54  & 9.03(-7)  \\
    31.36 & 2.45  & 0.54  & 9.03(-7)  \\
    34.46 & 2.45  & 0.54  & 9.03(-7)  \\
    37.66 & 2.45  & 0.54  & 9.03(-7)  \\
    39.26 & 2.45  & 0.54  & 9.03(-7)  \\
    41.06 & 2.45  & 0.54  & 9.03(-7)  \\
    42.86 & 2.45  & 0.54  & 9.03(-7)  \\
    44.56 & 2.45  & 0.54  & 9.03(-7)  \\
    46.41 & 2.45  & 0.54  & 9.03(-7)  \\
    47.76 & 2.45  & 0.54  & 9.03(-7)  \\
    50.81 & 2.45  & 0.54  & 9.03(-7)  \\
    52.21 & 2.45  & 0.54  & 9.03(-7)  \\
    54.76 & 2.45  & 0.54  & 9.03(-7)  \\
    56.35 & 2.42  & 0.54  &  1.89(-6) \\
    59.29 & 2.58  & 0.53  &  3.40(-6) \\
    63.07 & 2.42  & 0.51  & 2.54(-6)  \\
    65.95 & 2.54  & 0.50  & 2.10(-6)  \\
    68.12 & 2.49  & 0.46  & 1.75(-6)  \\
    70.30 & 2.61  & 0.48  & 9.92(-7)  \\
    71.95 & 2.52  & 0.48  & 1.08(-6)  \\
    75.05 & 2.82  & 0.48  &  5.60(-6) \\
    77.06 & 1.92  & 0.48  &  8.01(-7) \\
    80.32 & 2.73  & 0.48  & 2.10(-6)  \\
    85.15 &  2.37 & 0.49  & 3.33(-6)  \\
    88.96 & 1.86  & 0.41  & 1.74(-6)  \\
    92.12 & 2.33  & 0.48  & 1.52(-6)  \\
    94.98 & 2.52  & 0.48  & 2.39(-6)  \\ 
    99.06 & 2.33 & 0.53 & 8.14(-7) \\
    101.02 & 2.23 & 0.52 & 7.35(-7)  \\
    103.17 & 2.14 & 0.51 &  6.64(-7) \\
    105.62 & 2.45 & 0.54 &  9.09(-7) \\
    107.11 & 2.23 & 0.52 &  7.35(-7) \\
    108.90 & 2.33 & 0.53 &  8.14(-7) \\ 
    111.15 & 2.33 & 0.53 &  8.14(-7) \\
    113.83 & 1.56 & 0.40 &  2.06(-7) \\
    116.17 & 1.56 & 0.40 &  2.06(-7) \\
    122.10 & 1.36 & 0.20 &  1.17(-7) \\
   \hline
   \end{tabular}
   \end{flushleft}
    \end{table*}
 \subsection{Model simulation of 14.5\,GHz light curve}
    The model-simulation results for the 14.5\,GHz light-curve are displayed 
    in Figure 6. As for the 37\,GHz light-curve, the modeled 14.5\,GHz
    light-curve is derived from the optical light-curve by selecting the 
    spectral index $\alpha$(15GHz,V-band) for each of the thirty-six 14.5\,GHz
    elementary flares, which are listed in Table 7.  The simulation results
    are separately described for three segements.
    \subsubsection{Segment I}
      Only three 14.5\,GHz data-points are available in this time-interval 
    (JD2450024--2450049), we cannot make any meaningful conclusion about 
     the model-fitting. However, two data-points are close to the peaks of the
     elementary flares at
    $t_{\rm{peak}}$=31.36 and 44.56 within uncertainty of $\sim$1\,day, showing
    the modeled light-curve close to the observed level of  14.5\,GHz flux 
     density. In this time-interval the spectral index 
    ${\alpha}_{15,\rm{v}}$ was assumed to be 0.54. 
    \subsubsection{Segment II}
     For the main part of the optical outburst, the 14.5\,GHz light-curve
    has been simulated in terms of the lighthouse model. 
    The four 14.5\,GHz  data-points are 
    reasonably well fitted (within uncertainty of 
    $\sim$1-2\,days) by the  modeled total flux density curve at 
   $t_{\rm{peak}}$=50.8, 56.35, 59.29 and 65.95. In addition, ten 8\,GHz 
   data-points in the time-interval between 2450050 and 2450089 are 
   close to the modeled 14.5\,GHz total flux density curve, indicating 
    the simultaneous variations at 14.5--8\,GHz and optical wavelengths.
    \subsubsection{Segment III}
     Although only three data-points are available in this time-interval
    (JD2450090--2450118), two data-points are 
    reasonably well fitted by the modeled total flux curve at 
    $t_{\rm{peak}}$=103.17 and 105.62 within uncertainty of $\sim$1\,day. 
    The 14.5\,GHz data-point near JD2450097 might indicate the existence of 
    a radio burst not associated with any optical flare.\\
    \subsection{A brief summary}
    We have investigated the correlation between the optical flares and
    radio/mm bursts on timescales of days in OJ287 for the 1995.84
    optical outburst event. The aim is to search the clues for the lighthouse
    effect which produces the simultaneous optical and radio/mm elementary 
    flares with symmetric profiles. Obviously, the observation data 
    currently available at 37, 22, 14.5 and 8\,GHz are too sparse 
    to fulfill this task. However, we find that
    at least for the main part of the optical outburst between JD2450050 
     and 2450090, the modeled total 
    flux curves at 37\,GHz and 14.5\,GHz wavelengths reveal variations on 
    short time-scales which can reasonably well fit the observed radio
    peaks associated with  the optical peaks. Thus the lighthouse effect in
    producing the optical and radio/mm flares during the 1995.8 outburst 
    event has been confirmed. In order to further explore the correlation 
    between the optical and radio/mm variations on short time-scales 
    in OJ287, more contemporary observations at optical and radio/mm
    wavelengths on daily sampling rates are required.\\
     We further point out that the superluminal  optical knots producing the
    simultaneous optical and radio/mm flares near the nucleus 
      should continue to move outward along their trajectories and
     evolve to form superluminal radio knots on parsec scales, which
    emit at radio wavelengths, appearing as delayed radio flares.
   \begin{figure*}
   \centering
   \includegraphics[width=8cm,angle=-90]{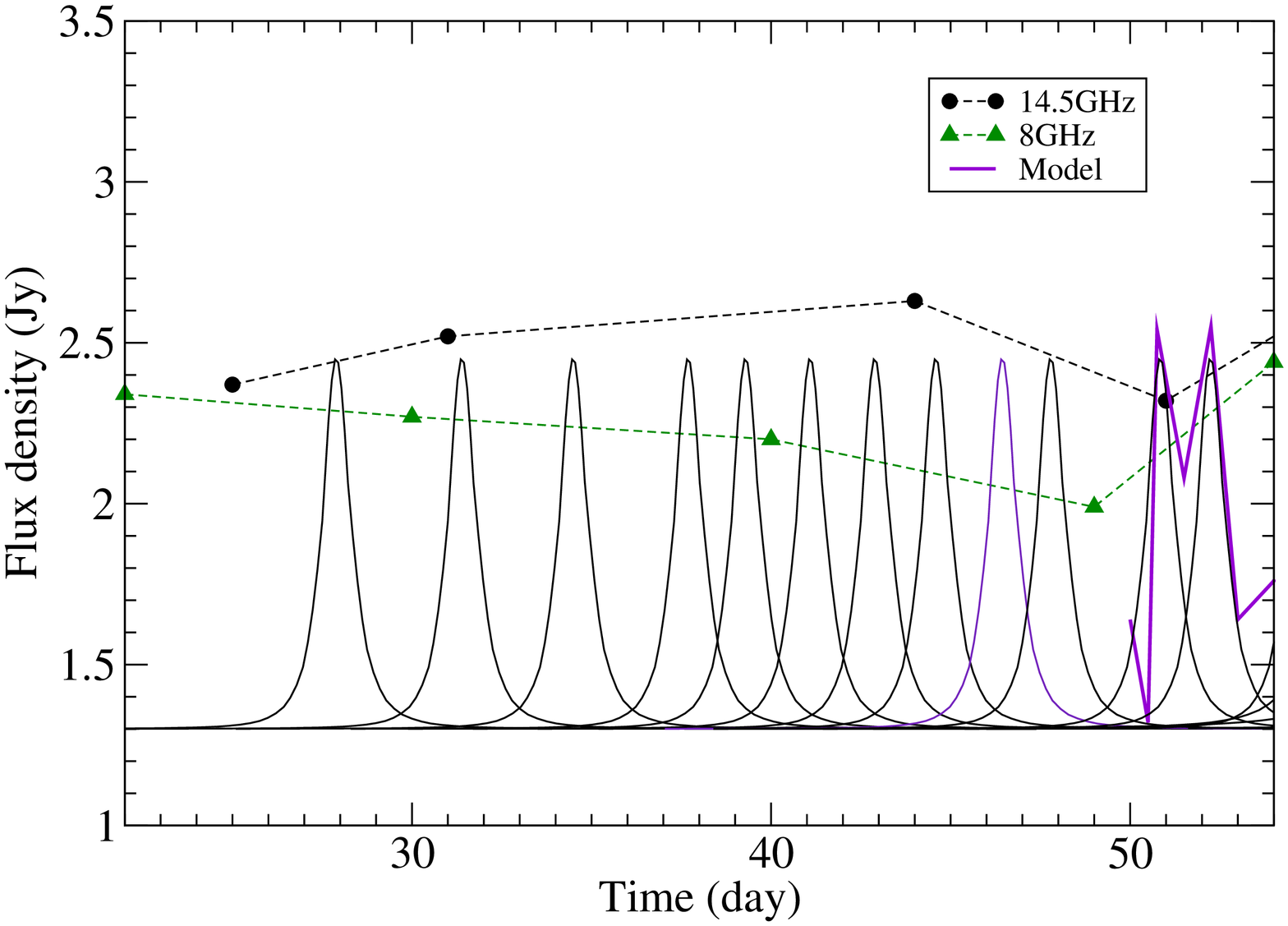}
   \includegraphics[width=8cm,angle=-90]{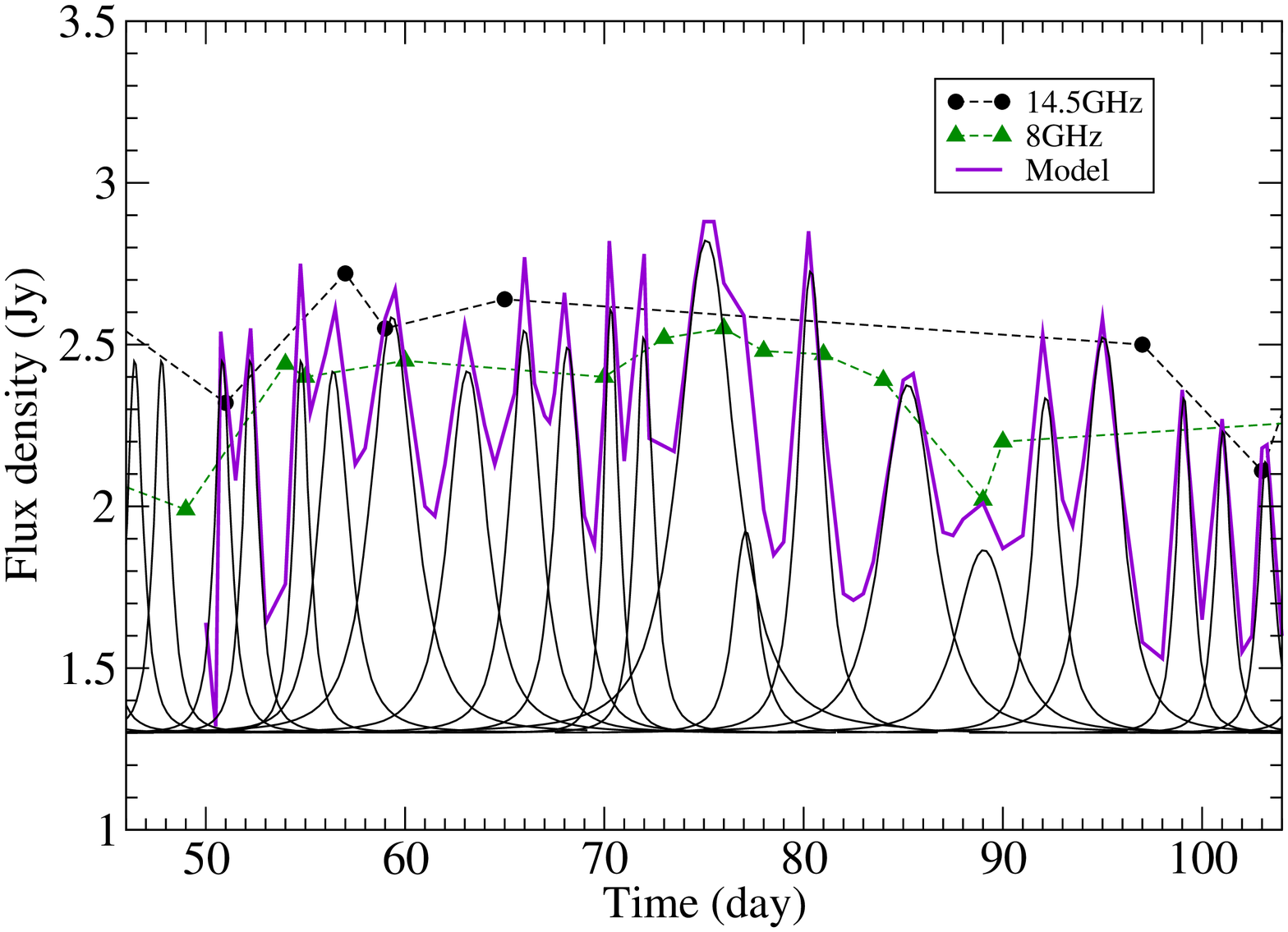}
   \includegraphics[width=8cm,angle=-90]{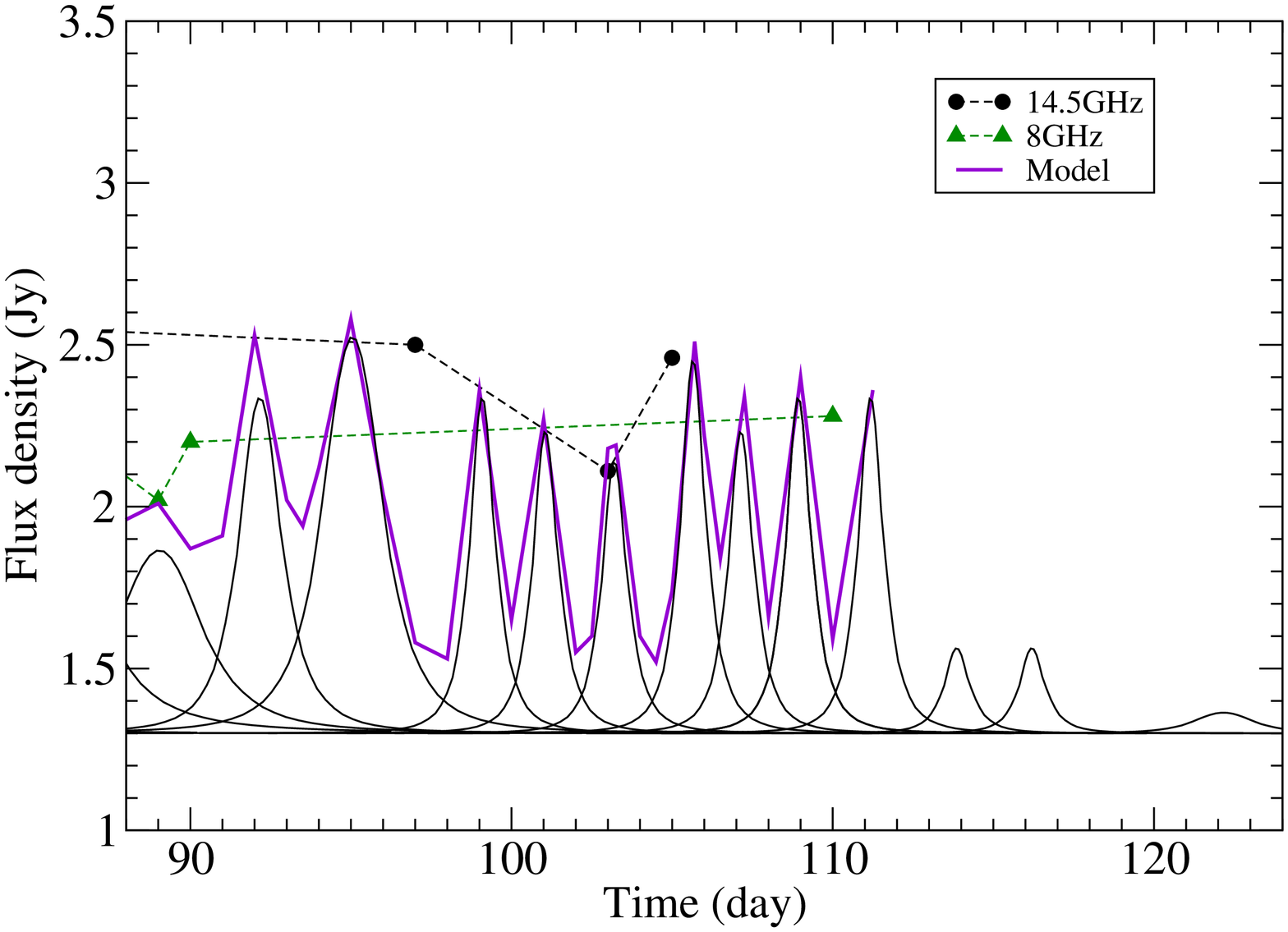}
   \caption{Model simulation of the 14.5\,GHz light curve. The model parameters
    are listed in Table 7.}
  \end{figure*}
   \section{Discussion}
    We have applied the precessing jet nozzle scenario previously proposed by
   Qian et al. (e.g., \cite{Qi91}, \cite{Qi13}, \cite{QiXiv19})
   to investigate and simulate the correlation between the variations 
  at optical and 
   radio wavelengths for the quasi-periodic optical outburst event during 
    1995.84--1996.10. Since  the radio/mm data are very sparsely sampled, this
   work can only be regarded as a preliminary examination. However, we have
   found convincing evidence that  the optical outburst can be decomposed 
   into a group of elementary flares and each of the optical elementary 
   flares emits simultaneous radio/mm emission. The simultaneous optical 
   and radio/mm variations can be reasonably well 
   interpreted in terms of lighthouse effect due to the helical motion 
   of the discrete superluminal optical/radio knots.\\
    Combining with the results obtained in Usher (\cite{Us79}), 
   Holmes et al. (\cite{Holm84}), D'Arcangelo et al. (\cite{Da09}) and 
   Qian (\cite{QiXiv19}), we may  tentatively 
   suggest that the quasi-periodic double-peaked outbursts in 1971--1973, 
   1983--1984, 1994--1995, 2005--2006 and the outburst in 2015.76 are 
   synchrotron flares produced within its relativistic jets 
   in OJ287. The superluminal optical/radio knots may have some
    kind of stratified core-envelope structure 
   \footnote{Or shocks of spine-sheath structure as suggested in 
   D'Arcangelo et al. (\cite{Da09}).} with its core region
     dominating the optical
   emission and the outer regions dominating the mm/radio emission. Thus 
   the optical and radio emission from the superluminal knots can be 
   Doppler boosted simultaneously due to lighthouse effect during their 
   helical motion.\\
   Helical motion of superluminal knots may be expected on the basis of 
   magnetodynamic (MHD) theories for jet formation in spinning black-hole and
   accretion-disk systems, in which relativistic jets are produced  in the 
   rotating magnetospheres with strong toroidal magnetic fields. Thus helical
   fields should permeate in the inner jet regions near the black holes (e.g.,
   Blandford \& Znajek \cite{Bl77}, Blandford \& Payne \cite{Bl82}, Camenzind
   \cite{Cam90}, Li et al. \cite{Lizy92}, Beskin \cite{Be10}, Vlahakis \&
    K\"onigl \cite{Vl04}, Meier \cite{Mei13}, \cite{Mei01}). It would be a 
    natural phenomenon that superluminal optical knots move along helical
    trajectories, producing optical outbursts through lighthouse effect.
    Unfortunately,  only a few optical events have been observed revealing
    this phenomenon (e.g., Schramm et al. \cite{Sc93}, 
    Dreissigacker \cite{Dr96a}, Dreissigacker \& Camenzind \cite{Dr96b},
    Camenzind \& Kronkberger \cite{Cam92}, Wagner et al. \cite{Wa95}). \\
    This work further strengthens the arguments proposed
    in Qian (\cite{QiXiv19} that helical motion of superluminal optical knots
    may prevail in blazar OJ287 and produce both quasi-periodic and 
    non-periodic optical outbursts, providing certain observational evidence
    for the existence of helical magnetic fields in its inner jet regions.\\
    Relativistic jet models (e.g., proposed in Qian \cite{QiXiv19} and 
   this work) can fully explain the emission properties from 
   radio, infrared, optical through to $\gamma$-rays observed in OJ287 and 
    a comprehensive and compatible framework has been tentatively constructed
    for interpreting the whole phenomena in blazar OJ287, which involve
    the following physical processes:
    (1) OJ287 may host a supermassive black hole binary with comparable
    masses for the primary and secondary holes in nearly co-planar eccentric
    orbital motion with an orbital period of $\sim$9\,yr (in the source frame);
   (2) Both black holes may be spining (Kerr holes) and create their 
    respective jets, directing toward the observer  with small viewing angles;
   (3) Cavity-accretion processes may occur in this black 
    hole/disk system due to the interaction between the spinning black-holes,
     the circumbinary disk and the circumdisks, causing complex mass
    accretion events;
   (4) The eccentric orbital motion of the binary modulates the 
    accretion flows onto the holes, resulting
   in the quasi-periodicity of the accretion events and consequential optical
    outbursts. The double-stream accretion flows during pericenter passages may
    result in the double structure of the quasi-periodic optical outbursts;
    (5) The accretion events created during the orbital motion would lead to 
   the ejection of superluminal optical knots through jet-formation mechanisms, 
   which emit at optical and radio wavelengths, including
   quasi-periodic outbursts and non-periodic intervening outbursts.\\
       In brief, it may be possible to describe the chain of the
    physical processes as follows. The interaction between the binary holes
   and the circumbinary disk results in double-stream accretion flows (e.g.,
   Tanaka \cite{Tan13}, Artymovicz \cite{Ar98}, Shi et al. \cite{Sh12},
   D'Orazio et al. \cite{Do13}) toward
   the binary holes during pericenter passages, creating superluminal optical
    knots ejected from the jets which produce the observed quasi-periodic 
   outbursts with double-peak structure. During the remaining parts of the
    orbit the interaction between the holes and circumbinary disk also creates
   enhanced accretion events, causing the non-periodic outbursts. The optical
   outbursts consist of elementary flares, which are produced by a 
   succession of individual superluminal optical knots due to 
   lighthouse effect and blend together to form the complex outbursts.\\
   If the relativistic jet scenario suggested in this work and 
    Qian (\cite{QiXiv19})
    is correct, the optical phenomena in OJ287 coul be explained without 
   needing to invoke the disk-impact mechanism and thermal impact-outbursts. 
   However, the disk-impact mechanism looks
    very attractive for testing the effects of general relativity
   (Einstein \cite{Ei16}, \cite{Ei18}), if the claimed accurate timing of the 
   primary quasi-periodic optical outbursts with uncertainty of $\sim$1\,day is 
   really confirmed in future.\\ 
    In order to test the relativistic models, more observational and 
   theoretical studies are needed. For finally clarifying the nature of optical 
   radiation from the primary quasi-periodic outbursts (thermal or 
   non-thermal),  observations of $\gamma$-rays associated with the primary
   quasi-periodic  optical outbursts are 
   important. Complete information about the time-variable polarization 
   (both polarization degree and polarization angle) of the primary
   quasi-periodic outbursts at multi-frequencies 
   are needed. Model fittings are also needed to 
   interpret the time-variable flux density curve I($\nu,t$), polarization
    degree curve p($\nu,t$) and polarization angle curve PA($\nu$,t).
    Generally, if the low polarization dip observed  in polarization 
   degree curves from  the primary quasi-periodic flares is associated with 
   the rapid change in its polarization angle, this quasi-periodic flare
    may be recognized as a synchrotron flare, originated from the relativistic
    jet (see Holmes et al. \cite{Holm84}). Observational studies of the
     correlation between variations in optical and radio 
    frequencies on time-scales of $\sim$1\,day will be very helpful, because
    radio variations simultaneous with the quasi-periodic optical outbursts
     will certainly imply the optical outbursts being non-thermal. \\
    For the relativistic models currently available, an unresolved problem
    may be related to the timing of the primary quasi-periodic optical 
    outbursts. In principle, cavity-accretion
    processes in comparable-mass and eccentric binary systems may provide
    some solutions to the quasi-periodicity and double-peak structure of the
    primary optical outbursts (Artymovicz \cite{Ar98}, Tanaka \cite{Tan13},
    Shi et al. \cite{Sh12}, D'Orazio et al. \cite{Do13} and 
    references therein). But theoretical model-simulations of the cavity 
    accretion processes for specifically 
    fitting the primary quasi-periodic optical outbursts and their
    double-structure pattern are required.  Non-periodic optical outbursts
    should also be included. It seems that in relativistic jet models
    the quasi-periodic optical outbursts can not be accurately timed  
    as required in the disk-impact model. In fact, 
    for the optical outburst during 1995.8--1996.1 we can only recognize
    the large synchrotron outburst during JD2450050--2450105 (see Figure 3).
    The problem of outburst timing for OJ287 needs to be further investigated.
    \begin{acknowledgement}
   I wish to thank Dr. M.~Aller (University of Michigan, USA) for affording 
    the 14.5 and 8\,GHz data on the optical outburst in 1995.84.
    I am most grateful to 
   Dr. K.~Nillson (Turku University Observatory, Finland) for providing the 
   optical data during the OJ-94 project.
   \end{acknowledgement}

  \end{document}